\documentclass[]{jfm}
\usepackage{graphicx}
\usepackage{epstopdf,epsfig}
\usepackage{fix-cm}
\usepackage{natbib}
\usepackage{hyperref}
\usepackage{natbib}
\usepackage{amsmath,mathtools}
\usepackage{longtable} 
\usepackage{tikz}
\usetikzlibrary{patterns}
\usepackage{adjustbox}
\usepackage{color}
\hypersetup{
    colorlinks = true,
    urlcolor   = blue,
    citecolor  = black, 
}
\usepackage{bm}
\usepackage{subcaption} 
\usepackage{enumitem} 
\definecolor{caribbeangreen}{rgb}{0.0, 0.8, 0.6}

\newcommand{\RomanNumeralCaps}[1]
\linenumbers

\newcommand{\beq}{\begin{equation}}
\newcommand{\eeq}{\end{equation}}
\newcommand{\per}{\, .}
\newcommand{\com}{\, ,}

\newcommand{\half}{\tfrac{1}{2}}

\newcommand{\Nu}{\mbox{\textit{Nu}}}
\newcommand{\Nuchi}{\Nu^{ \chi}}
\newcommand{\Nus}{\Nu_{\mathrm s}}
\newcommand{\NuF}{\Nu_{\mathrm F}}

\newcommand{\Ra}{\mbox{\textit{Ra}}}
\renewcommand{\Re}{\mbox{\textit{Re}}}
\newcommand{\RaH}{\mathrm{Ra_H}}
\newcommand{\RaHtrad}{\mathrm{Ra^{trad}_H}}
\newcommand{\RaF}{\textcolor{red}{\mathrm{Ra_V}}}
\newcommand{\RaV}{\mathrm{Ra_V}}

\newcommand{\RaHN}{\mathrm{Ra}^{\mathrm{N}}_\mathrm{H}}
\newcommand{\RaHstrg}{\mathrm{Ra}^{\mathrm{S}}_\mathrm{H}}
\newcommand{\hatRaH}{\mathrm{\hat{R}a_H}}
\newcommand{\hatRaF}{\mathrm{\hat{R}a_V}}

\renewcommand{\Pr}{\mbox{\textit{Pr}}}

\def\P{\mathrm{P}}
\def\Q{\mathrm{Q}}
\def\R{\mathrm{R}}

\newcommand{\ep}{\epsilon}

\newcommand{\UI}{U_\text{I}}

\renewcommand{\S}{\mathscr{S}}

\newcommand{\Smax}{\mathscr{S}_{\mathrm{max}}}
\newcommand{\Vmax}{\mathscr{V}_{\mathrm{max}}}
\newcommand{\Tmax}{\mathscr{T}_{\mathrm{max}}}
\newcommand{\Wmax}{\mathscr{W}_{\mathrm{max}}}

\newcommand{\psh}{\hat \psi}
\newcommand{\thh}{\hat \theta}
\newcommand{\vth}{\vartheta}
\newcommand{\alp}{\alpha}
\newcommand{\tha}{\theta}
\newcommand{\be}{\beta}

\newcommand{\da}{\delta}
\newcommand{\dab}{\delta_b}
\newcommand{\dau}{\delta_u}
\newcommand{\daR}{\delta_{\mathrm{R}}}

\newcommand{\bu}{\boldsymbol{u}}
\newcommand{\grad}{\boldsymbol{\nabla}}
\newcommand{\bx}{\boldsymbol{x}}
\newcommand{\lap}{\nabla^2}

\newcommand{\ellx}{\ell_x}
\newcommand{\elly}{\ell_y}
\newcommand{\bs}{b^{\text{s}}}
\newcommand{\bstar}{b_{\star}}
\newcommand{\Fb}{F_b}

\newcommand{\defn}{\ensuremath{\stackrel{\mathrm{def}}{=}}}

\newcommand{\la}{\langle}
\newcommand{\laa}{\left\langle}
\newcommand{\ra}{\rangle}
\newcommand{\raa}{\right\rangle}

\newcommand{\bdiff}{c}
\newcommand{\chidiff}{\chi^{c}}

\newcommand{\ga}{\gamma}

\newcommand\WRYcom[1]{\textcolor{magenta}{[[WRY: #1]]}}
\newcommand\WRY[1]{\textcolor{magenta}{#1}}
\newcommand\SGLS[1]{\textcolor{blue}{#1}}
\newcommand\FR[1]{\textcolor{purple}{#1}}
\newcommand{\restrat}{\textcolor{magenta}{restratification}}

\newcommand\SR[1]{\textcolor{red}{\sout{#1}}}

\title[Restratification by horizontal convection]{Suppression of Rayleigh-B\'enard convection and restratification by horizontal convection}
\shortauthor{F. Rein, S. G. Llewellyn Smith and W. R. Young}

\author{Florian Rein\aff{1} \corresp{\email{florian.rein@protonmail.com}}, 
Stefan G. Llewellyn Smith\aff{1,2} 
  \and
  William R. Young\aff{1}}

\affiliation{ \aff{1} Scripps Institution of Oceanography, University of California San Diego, La Jolla CA 92093-0213, USA
\aff{2} Department of Mechanical and Aerospace Engineering, Jacobs School of Engineering, University of California San Diego, La Jolla CA 92093-0411, USA
}

\begin{document}
\maketitle

\begin{abstract}
We investigate the competition between horizontal convection (HC) and Rayleigh–B\'enard convection (RBC) in a fluid layer subject to a uniform destabilizing  buoyancy flux at the bottom and a horizontally varying buoyancy distribution at the top. The RBC forcing imposes negative horizontal mean vertical buoyancy gradients at the top and bottom of the fluid layer. But if the HC forcing is sufficiently strong then  the volume averaged vertical buoyancy gradient, $\langle b_z \rangle$, is positive i.e.~opposite in sign to destabilizing RBC buoyancy  gradients at the boundaries. If $\langle b_z \rangle>0$ we say that the layer has been ``restratified''.

Using scaling analysis based on power integrals together with two-dimensional direct numerical simulations at Rayleigh numbers up to $10^{10}$, we identify two cases: a neutral stratification state, in which HC first offsets RBC so  that $\langle b_z \rangle = 0$, and a strong stratification regime, in which HC dominates and $\langle b_z \rangle$ is opposite in sign, and greater in magnitude, than the prescribed  destabilizing vertical buoyancy gradient at the layer boundaries. For the range of parameters explored in this study, we derive scaling laws for the onset  of these regimes in terms of the horizontal and vertical flux Rayleigh numbers, $\RaH$ and $\RaV$, finding $\RaHN \sim \RaV^{4/5}$ for the neutral state and $\RaHstrg \sim \RaV$ for the onset of strong stratification. The results highlight the controlling role of the top boundary layer in setting the mean stratification and clarify the conditions under which HC suppresses RBC. 

These findings are relevant to geophysical environments such as subglacial lakes, and the oceans of Snowball Earth and icy moons, where bottom heating and horizontal buoyancy variations jointly shape ocean stratification.
\end{abstract}

\begin{keywords}
Rayleigh-B\'enard convection, horizontal convection, geothermal heat flux, ice-covered satellites, Neoproterozic Earth
\end{keywords}
%

\section{Introduction \label{intro}}

Geothermal heating of the Earth's oceans has a flux Rayleigh number of order  $10^{24}$ (see \autoref{fluxRa}). This strongly supercritical  bottom heating does not result in  Rayleigh-B\'enard convection (RBC). Instead,  ocean stratification is reliably  stable --  even strongly stable in the sense that  the buoyancy frequency is greater than the Coriolis frequency almost everywhere. Suppression of geothermally forced  RBC indicates the importance of horizontal convection \citep{rossby1965,rossby1998,ANRHC2008} associated with  a sea-surface pole-to-equator temperature difference of order  $30$ K. Horizontal convection (HC), assisted by  wind and tides,  produces a stably stratified ocean \citep{munk1998}. Impelled by  observations of stable ocean stratification, oceanographers  are   concerned only with secondary effects of geothermal heating,  such as modification of abyssal water masses  \citep[e.g.][]{emile2009,deLavergne2016} or the possibility that geothermal heating enhances  the strength of the meridional overturning circulation \citep*[e.g.][]{mullarney2006,wang2016}. 

\begin{figure}
  \centering
  \includegraphics[width=0.8\textwidth]{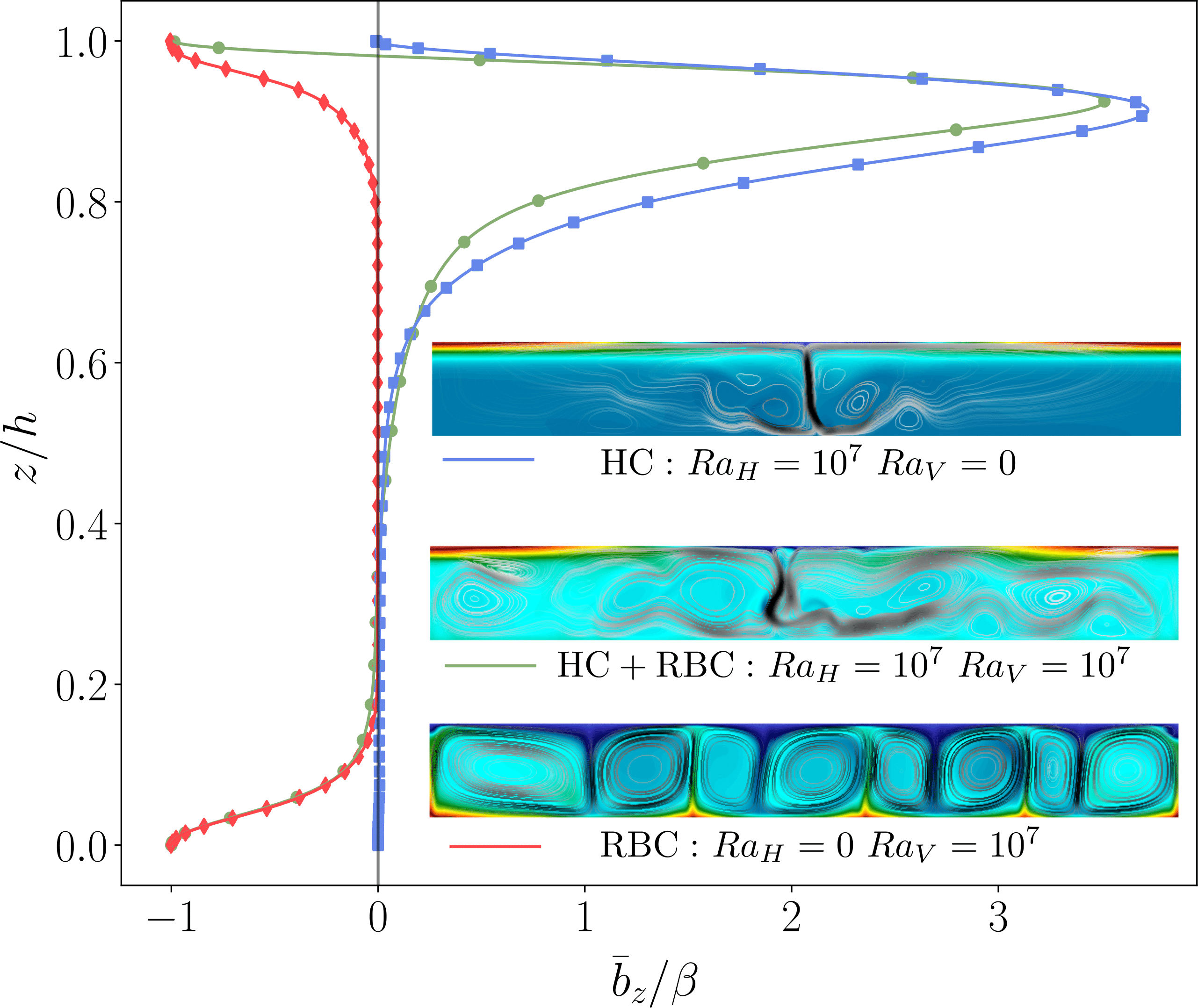}
    \caption{Horizontally- and time-averaged vertical buoyancy gradient $\bar b_z(z)$ in three cases. The inset shows a snapshots of the buoyancy with overlaid streamlines. The HC Rayleigh number, $\RaH$, and the RBC flux Rayleigh number, $\RaV$, are defined in \eqref{RaZ}; other parameters are $\Gamma =8$ and $\Pr=1$. For the joint HC and RBC case, the time and volume-averaged vertical buoyancy gradient is $\langle b_z \rangle/\beta=0.38$, where $\beta>0$  is defined in \eqref{FBC}. }
  \label{fig1}
\end{figure}

We refer to the  production of stable interior stratification in  a bottom-heated  fluid  layer  as  ``restratification.''  As a  definition, we say that the fluid is  restratified if the volume average of the vertical buoyancy gradient -- $\la b_z\ra$ in the notation of section~\ref{form} -- is positive. \autoref{fig1} shows an example of restratification: in the HC+RBC configuration the volume-averaged vertical buoyancy gradient $\la b_z\ra$ is positive, despite bottom heating.


Geophysical convection problems   involving restratification  include the circulation of subglacial lakes \citep{THOMA2009,couston2021}, the ocean of snowball Earth  during the Neoproterozoic \citep{hoffman2002,pierrehumbert2011} and the oceans of icy moons such as Europa and Enceladus \citep*[e.g.][]{soderlund2019,Lemasquerier2023}. In these systems geothermal  heating is essential in maintaining a liquid ocean beneath an ice sheet. The ice sheet has variable thickness, and therefore variable pressure and  freezing temperature at its base. The non-uniform basal  ice temperature results in HC, and perhaps restratification, in the geothermally heated water below.  Other horizontal non-uniformities result from spatial variations in the bottom flux of heat, salinity fluxes from the freezing and melting of ice, and  non-uniform depth of the water layer. 

The oceans of icy moons and of snowball Earth  differ from Earth's present ocean in that there is no  sea-surface forcing by wind stress. Wind stress is the main source of energy for ocean circulation \citep{munk1998,wunsch2004}. It is likely therefore that the circulation of icy-moon oceans differs qualitatively from that of Earth: observations of ubiquitous statically stable  stratification in Earth's present ocean are not a reliable guide to  stratification in other bottom-heated oceans.

Investigations of snowball Earth and icy moons make diverse and contradictory assumptions regarding the competition between RBC and HC in these systems. For example, \cite{jansen2016} assumes that horizontal inhomogeneities in snowball Earth  heat fluxes result in statically stable stratification. In Jansen's view, vertical  heat fluxes, required to transmit geothermal heat to the base of the ice sheet,  result from  baroclinic instability (also known as slantwise convection) in a horizontally restratified ocean. On the other hand, \cite{ashkenazy2016} argue that the ocean of snowball Earth is, to a first approximation,  well mixed vertically by geothermal RBC.  In a numerical exploration of icy-moon ocean circulation, \cite{bire2022} ignore  horizontal inhomogeneities that might result in HC and focus instead on rotating RBC forced by unopposed uniform bottom heating. But in  another exploration of icy-moon ocean circulation, \cite{zhang2024ocean} use an insulating bottom boundary condition (no RBC) and instead force with horizontally non-uniform temperature (i.e.~HC) at the base of the ice sheet.

Models of icy-moon  oceans are largely  unconstrained by observations. For snowball Earth, ignorance is even more profound and likely to remain so. It is essential therefore to identify the parameters determining whether the bulk stratification $\la b_z\ra$ of these oceans is statically stable. Subsequent modeling  depends heavily on this issue. For example, a stably stratified ice-covered ocean might be energized by tidally generated internal waves resulting in turbulent interior mixing \citep[e.g.][]{Wunsch2016}. Baroclinic instability  occurs only if the bulk stratification is   stable. Sufficiently strong stable stratification is the basis for the ``traditional approximation'' in which the component of the Coriolis force along the direction of the base-state buoyancy gradient is neglected \citep[e.g.][]{gerkema2008}. In short, if $\la b_z\ra >0$ then business-as-usual ocean modeling can proceed. If $\la b_z\ra < 0$ then a very different class of models, suggested by processes in a stellar convection zone, might be appropriate.

For both RBC and HC a key challenge is determining the relation between  heat transport  and  surface forcing, i.e.~the relation between Nusselt $\Nu$ and Rayleigh $\Ra$ numbers. While the $\Nu$-$\Ra$ relation is not a goal of this investigation,  related boundary-layer scaling arguments are useful in section \ref{phenomenology}. For the RBC boundary layer, see \cite{GL2000,GL2001}, \cite{doering2020turning} and \cite{lindborg2023scaling}. For the HC boundary layer the literature is smaller: see  \cite{rossby1965,rossby1998}, \cite{Shishkina2016} and more recently \cite{Passaggia_Scotti_2024} and \cite{Passaggia_Cohen_Scotti_2024}.

Motivated by the dynamics of subglacial lakes, \cite*{couston2022} (hereafter CNF) investigated the competition between RBC and HC and  demonstrated a regime transition between the two varieties of convection. Close to the regime transition, and  with flux Rayleigh number as large as $10^6$,  CNF showed that  there is bistability and hysteresis:  two stable flow states co-exist at the same parameter values. In one state RBC is dominant (multiple convection cells, most with a horizontal scale comparable to the layer depth)  and in  the other state HC dominates (fewer cells, most elongated in  the horizontal direction). The history of the system determines which state is observed.

Here we use the RBC+HC  setup studied by CNF, focusing on the issue of restratification.   Restratification  indicates that the competition between RBC and HC is settled: RBC has capitulated and most of the  fluid is statically stable with $\la b_z\ra$ having the positive sign characteristic of HC.

Geophysical situations present  complications such as rotation, salinity, and spherical shell geometry. But understanding restratification in the idealized RBC+HC  model is a necessary prerequisite: we must first identify the non-dimensional  parameters that determine the outcome of this competition in the relatively simple RBC+HC model.


Section~2 presents the governing equations, numerical setup, and control parameters. We define the two Rayleigh numbers that quantify the bottom buoyancy flux and the horizontal buoyancy forcing.
Section~3 defines restratification, and section~4 examines the flow phenomenology and key scalings of two characteristic stratification configurations: neutral and strong stratification.
Section~5 develops scaling arguments based on the dynamics of the top boundary layer and identifies its controlling role in the system.
We derive scaling laws for the boundary-layer thickness and mean buoyancy in the asymptotic regimes in which RBC dominates horizontal convection and vice versa.
The same framework is then used to determine the thresholds for the onset of neutral and strong stratification, establishing their dependence on the flux and horizontal Rayleigh numbers in both low- and high-Rayleigh regimes.
Section~6 examines the sensitivity of the results to variations in the Prandtl number, compares them with the findings of CNF, and explores the implications for geophysical systems where bottom heating and lateral buoyancy gradients interact.
Section~7 concludes with directions for future work.

\section{Formulation of the restratification problem \label{form}}
\subsection{The Boussinesq equations}
We consider a layer of Boussinesq fluid with depth $h$ and density $\rho = \rho_{\text{ref}} (1 - g^{-1}b)$, where $\rho_{\text{ref}}$ is a constant reference density, $g$ is the gravitational acceleration and $b$ is the ``buoyancy''.
If, for example, the fluid is stratified by temperature variations then $b = g \alpha (T - T_\text{ref})$, where $T_\text{ref}$ is a reference temperature, $\alpha$ is the thermal expansion coefficient and $g$ is the gravitational acceleration.We define $T_\text{ref}$ so that the horizontal and time average of  $b$ prescribed at the top surface $z=h$ is zero.

At the bottom, $z = 0$, the buoyancy boundary condition is  specified constant  diffusive flux of buoyancy, $F$ defined in \eqref{gT19}, into the fluid. This is equivalent to the Neumann boundary condition
\beq
 b_z(x,y,0,t) = - \beta \com
\label{FBC}
\eeq
where  $\beta = F/\kappa>0$.
At the top of the layer, $z = h$, the boundary condition is 
\beq
b(x,y,h,t) = \bs(x,y)\com
  \label{sbuoy8}
\eeq
where  the ``surface buoyancy'' $\bs$ is a specified function of position, e.g.~\eqref{sbuoy17} below. 


\input{figures/fig2}

With the notation above, the Boussinesq equations of motion are
\begin{align}
  \bu_t + \bu \bcdot \grad \bu + \bnabla p &= b \boldsymbol{\hat z} + \nu \lap \bu \com \label{mom} \\
  b_t + \bu \bcdot \grad b &= \kappa \lap b \com \label{buoy} \\
  \bnabla \bcdot \bu &= 0 \per \label{divu}
\end{align}
The kinematic viscosity is $\nu$ and the thermal diffusivity is $\kappa$.
We impose no-slip boundary conditions on all solid boundaries.
In two-dimensional  numerical solutions,  with $-\ellx/2 < x < + \ellx/2$, we use the sinusoidal profile
\beq
  \bs(x) = - \bstar \cos k x \com
  \label{sbuoy17}
\eeq
with $k = 2 \pi / \ellx$.
In general one can introduce a buoyancy scale $\bstar$ as a dimensional measure of horizontal buoyancy variations specified  at the top surface.
We take $\bstar>0$ so that the densest surface fluid is in the middle of the domain at $x=0$. The HC plume  is then well away from the no-slip sidewalls at $x=\pm\ellx/2$ (see inset in figure \ref{fig1}).
This setup is  illustrated in \autoref{fig2}. 

Numerical solutions of \eqref{FBC} through \eqref{sbuoy17} presented in section  \ref{phenomenology} and \ref{sec:scalings} are based on the two-dimensional (2D) case with $\elly=0$. The theoretical developments in section \ref{defn}, however,  applies equally well to the three-dimensional (3D) case. And in section \ref{discussion} we assess the effects of  three-dimensionality by presenting numerical solutions with periodicity in the $y$-direction and $\ell_y=h$.

\subsection{Non-dimensional control parameters}

This problem is characterized by both a HC Rayleigh number and a vertical RB flux Rayleigh  number:
\beq
  \RaH \defn \frac{h^3 \bstar}{\nu \kappa} \qquad \text{and} \qquad \RaV \defn \frac{h^4F}{\nu\kappa^2} =   \frac{h^4\beta}{\nu\kappa}\per
  \label{RaZ}
\eeq 
Additional non-dimensional parameters are the Prandtl number and aspect ratio:
\beq
\Pr \defn \frac{\nu}{\kappa} \qquad \text{and} \qquad  \Gamma \defn \frac{\ellx}{h}\per
\eeq

The traditional definition of the HC Rayleigh number, introduced by \cite{rossby1965,rossby1998}, is based on the  horizontal length $\ellx$ and the total difference between the maximum and minimum imposed surface buoyancy. In our notation, and using the sinusoidal surface buoyancy  profile in \eqref{sbuoy17}, this traditional HC Rayleigh number is 
\beq
\RaHtrad =  \frac{2 \bstar \ellx^3}{\nu\kappa} = 2 \Gamma^3 \RaH\per
\label{RaTrad}
\eeq
Our unconventional definition of $\RaH$ simplifies subsequent results. But because we consider largish aspect ratios ($\Gamma\geq 8$)  our $\RaH$ is smaller than the corresponding $\RaHtrad$ by a factor of at least $2^{10}$. This factor $2^{10}$  complicates comparison of our results with earlier studies of HC  using $\RaHtrad$ as a control parameter.

We introduce non-dimensional variables denoted by a dash and defined by
\begin{gather}
(x,y,z) = h (x',y',z')\com \qquad t' = \kappa t/h^2 \label{ndvar7}\com\\
\bu' = h\bu/\kappa \com \qquad p' = h^2 p/(\nu\kappa) \com \qquad  b' = h^3b/(\nu \kappa) \per \label{ndvar77}
\end{gather}
Dropping the dash, the non-dimensional versions of \eqref{mom} through \eqref{divu} are
\begin{align}
  \Pr^{-1}(\bu_t + \bu \bcdot \grad \bu) &= -\bnabla p +b \boldsymbol{\hat z} +   \lap \bu \com \label{mom7} \\
  b_t + \bu \bcdot \grad b &=  \lap b \com \label{buoy7} \\
  \bnabla \bcdot \bu &= 0 \per \label{divu7}
\end{align}
The non-dimensional versions of the boundary conditions in \eqref{FBC} and \eqref{sbuoy17}  are 
\beq
  b(x,y,1,t) = - \RaH \cos k x \com \qquad b_z(x,y,0,t) = - \RaV\per
  \label{ndBCZ}
\eeq
The unconventional definition of non-dimensional variables in \eqref{ndvar7} and \eqref{ndvar77} ensures that   Rayleigh numbers $\RaH$ and $\RaV$ appear on equal footing in the boundary conditions \eqref{ndBCZ}. Using $\RaH$, rather than $\RaHtrad$, ensures that the aspect ratio $\Gamma$ does not appear in the boundary conditions \eqref{ndBCZ}.

The governing equations (\ref{mom7}--\ref{divu7}), together with the boundary conditions \eqref{ndBCZ}, and no-slip conditions on all boundaries, are solved numerically in a two-dimensional box using the spectral-element code \href{https://nek5000.mcs.anl.gov/}{Nek5000} \citep{Fischer1997,Deville2002}.
This solver has been widely applied to studies of thermal convection \cite[e.g.][]{Scheel2013,rein2023}.
Further details of the numerical method and its implementation are in \autoref{num_method}.

Pure RBC corresponds to  $\RaH=0$, or equivalently $\bstar=0$.  This is  type 3 RBC as defined by \cite{goluskin2016}: the lower boundary condition is fixed flux and the upper boundary condition is fixed uniform temperature. 
With $\bstar=0$ there is a motionless diffusive solution with $\bu = \bf{0}$ and $b=\beta(h-z)$. This solution is linearly unstable if  $\RaV>  1295.78$ and is globally stable if $\RaV < 1295.78$ \citep{goluskin2016}. Type 3 RBC is cellular: at onset the most unstable horizontal wavenumber is $kh = 2.5519$.

Pure HC corresponds to $\RaV=0$. In HC the smallest temperature variation at $z=h$ create horizontal pressure gradients that set the fluid into motion. i.e.~the critical value of $\RaH$ is zero. In contrast to cellular RBC, HC produces overturning flow with a horizontal length scale determined by the applied surface buoyancy; see ~\autoref{fig1}.

\section{Definition of restratification \label{defn}}

\subsection{Power integrals}

We use an overbar to denote an average over $x$, $y$ and~$t$, taken at any fixed $z$, e.g.~see $\bar b_z$ in figure~\ref{fig1}. Angle brackets $\la \; \ra$ denote a total  volume and time average. For \textcolor{red}{operational}  details of the averaging process, see Appendix~\ref{num_method}.

Forming  $\la \bu \bcdot~\eqref{mom} \ra$ produces the energy power integral
\begin{align}
  \varepsilon = \la w b \ra \com \label{PY2}
\end{align}
where $\varepsilon \defn \nu \la |\grad \bu|^2 \ra>0$ is the rate of dissipation of kinetic energy and $\la w b \ra$ is the rate of conversion from  internal and gravitational potential energy to  kinetic energy.

Averaging the buoyancy equation~\eqref{buoy} over the horizontal coordinates  and time $t$, and then integrating in $z$, leads to the flux constraint
\beq
  \overline{wb} - \kappa \bar b_z =\kappa \beta\per
  \label{PY1}
\eeq
In pure HC, with $\beta=0$, \eqref{PY1} is the zero-flux constraint of  \cite{paparella2002}. With non-zero buoyancy  flux through the bottom, \eqref{PY1} says that in steady state the same buoyancy flux $\kappa \beta$ must pass through every level.

Taking the $z$-integral of \eqref{PY1}, dividing by $h$, and substituting into \eqref{PY2}  gives
\begin{align}
 \varepsilon  =\kappa \beta +  \kappa \la b_z\ra  \per \label{veps2}
\end{align}
 In \eqref{veps2} $\varepsilon$ and $\kappa \beta$ are both positive. But the third term, $\kappa \la b_z\ra$, might have either sign. From $\la b \eqref{buoy}\ra$ we obtain the buoyancy power integral
\beq
\chi = \kappa  \overline{b_z(h) \bs(x)}/h - \kappa \beta \la b_z \ra\com
\label{buoyDiss}
\eeq
where $\chi \defn \kappa \la |\grad b|^2\ra$ is the dissipation of buoyancy variance and $b_z(h)= b_z(x,y,h,t)$ \citep[e.g.][]{rocha2020,winters2009}. Although $\chi$ is positive  the terms on the right of \eqref{buoyDiss} can have either sign.

\subsection{Neutral stratification, restratification,  and strong restratification}
In pure HC ($F=\beta \kappa =0$) it follows from \eqref{veps2} that $\la b_z\ra >0$.  The other limiting case is pure type 3 RBC, corresponding to $\bs(x)=0$ and $F >0$. In this case \eqref{buoyDiss} shows that $\la b_z\ra <0$. To summarize:
\begin{align}
\text{pure HC \eqref{veps2}} \qquad    &\Rightarrow  \qquad \la b_z \ra >0\,; \\
\text{pure RBC \eqref{buoyDiss}}  \qquad &\Rightarrow  \qquad \la b_z \ra < 0\per
\end{align}
The results above, based on power integrals, motivate using the sign change of $\la b_z \ra$ to define the onset of ``restatification'': we say that a bottom-heated layer with non-uniform surface temperature is restratified by HC if $\la b_z \ra > 0$. 

At the onset of restratification, with
\beq
\la b_z\ra = 0\com
\eeq 
we say that the layer has ``neutral stratification''. Recalling that $\bar b(h)=0$, the identity 
\begin{align}
\la b_z\ra &=  -\frac{ \bar b(0)}{h} \label{delb3}
\end{align}
shows that a neutrally stratified layer has  $\bar b(0)=0$.

From the HC extremum principle \citep{paparella2002}, the smallest (most negative) buoyancy in the domain must be greater than the smallest  buoyancy prescribed at the surface $z=h$. Hence for the sinusoidal surface profile in \eqref{sbuoy17} we obtain  from \eqref{delb3}
\begin{align}
\la b_z\ra \leq \frac{\bstar}{h} \per \label{delb13}
\end{align}
 Restating \eqref{delb13} in dimensionless variables produces an upper bound on the possible strength of restratification
\beq
\frac{\la b_z\ra}{\beta} \leq \frac{\RaH}{\RaV}\per
\label{delb17}
\eeq

As  a definition of the  situation in which  HC dominates RBC we say that the layer is ``strongly restratified'' if
\beq
1 \leq \frac{\la b_z \ra }{\beta}\com \qquad \text{(definition of strong restratification).}
\label{strong1}
\eeq
The upper bound in \eqref{delb17} shows that strong restratification is not possible if $\RaH/\RaV < 1$.

The bottom boundary condition is $b_z(x,y,0,t)=-\beta$, so strong restratification means that the bulk stratification, $\la b_z\ra$, is equal in magnitude, but opposite in sign, to   the destabilizing  bottom buoyancy gradient. This definition  of strong restratification is somewhat arbitrary. But solutions discussed in section~\ref{phenomenology} indicate that  with \eqref{strong1} the  top boundary layer   is characteristic of HC, i.e. if inequality \eqref{strong1} applies then RBC has capitulated to HC.

Rotation does not alter the power integrals in \eqref{veps2} and \eqref{buoyDiss}.
Thus the  definitions of neutral and strong stratification, and all results based on power integrals,  extend  to rotating   flows.

\section{Phenomenology and the surface boundary layer \label{phenomenology}}
We describe the flow characteristics associated with  neutral and strong restratification.  
\begin{figure}
  \centering
  \includegraphics[width=1.0\textwidth]{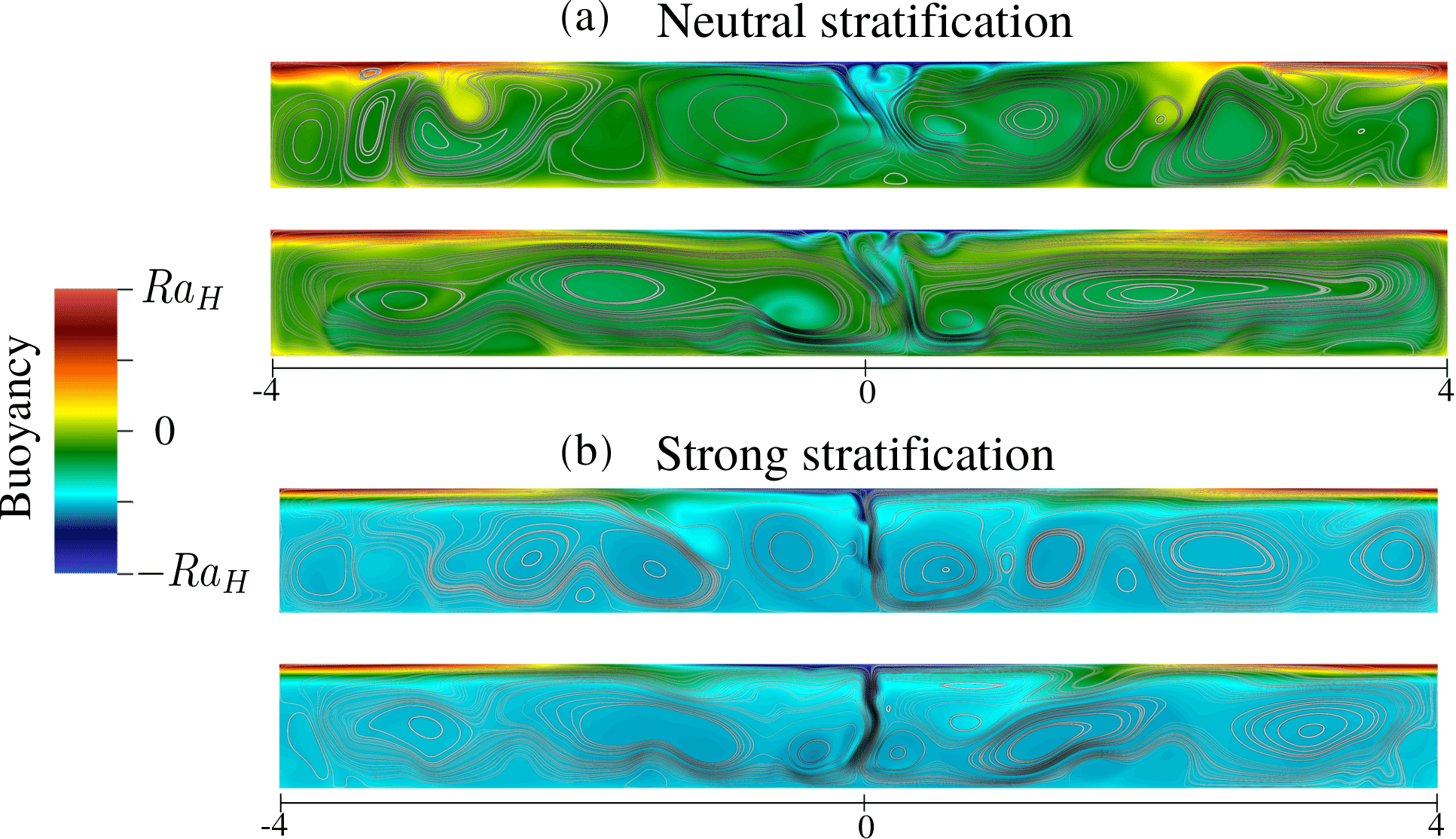}
    \caption{Two snapshots of the buoyancy field separated by a short time interval around $0.02\,h^2/\kappa$ for $(a)$ the neutral stratification state ($\RaH = 3.35\times10^6$) and $(b)$ the onset of the strong stratification regime ($\RaH = 2.03\times10^7$). Streamlines are overlaid in both cases. Parameters are $\RaV = 10^7$, $\Gamma = 8$, and $Pr = 1$. The buoyancy unit is $\nu\kappa/h^3$.}
  \label{fig3}
\end{figure}

\subsection{Neutral stratification}

In the neutrally stratified state, with $\la b_z\ra=0$, rising plumes generated by the bottom buoyancy flux are intermittently swept along the bottom towards the sidewalls by a large-scale horizontal circulation, as observed in CNF. This circulation is the bottom expression of the dominant HC overturning cell, which is in turn driven by the descending plume that forms near the centre of the upper boundary. The flow alternates between RB-like convection cells and the broader HC circulation  -- see \autoref{fig3}(a).

Profiles of  $\bar b(z)$  with neutral stratification are shown in \autoref{fig4}(a). The buoyancy is normalized by the imposed HC surface amplitude $b_*$, and the vertical coordinate is rescaled by the thickness of the top boundary layer $\delta$, defined as the distance from the top boundary at which $\bar b_z$ reaches $95\%$ of its maximum value.  When presented  in this way, $\bar b(z)$ profiles with  $\RaV \geq 10^6$ collapse onto a common curve in the top boundary layer. Thus with $\la b_z\ra=0$, the buoyancy variation across the top boundary layer 
is controlled by the imposed HC surface forcing $\bstar$.

\subsection{Strong stratification}

\begin{figure}
  \centering
  \includegraphics[width=1.0\textwidth]{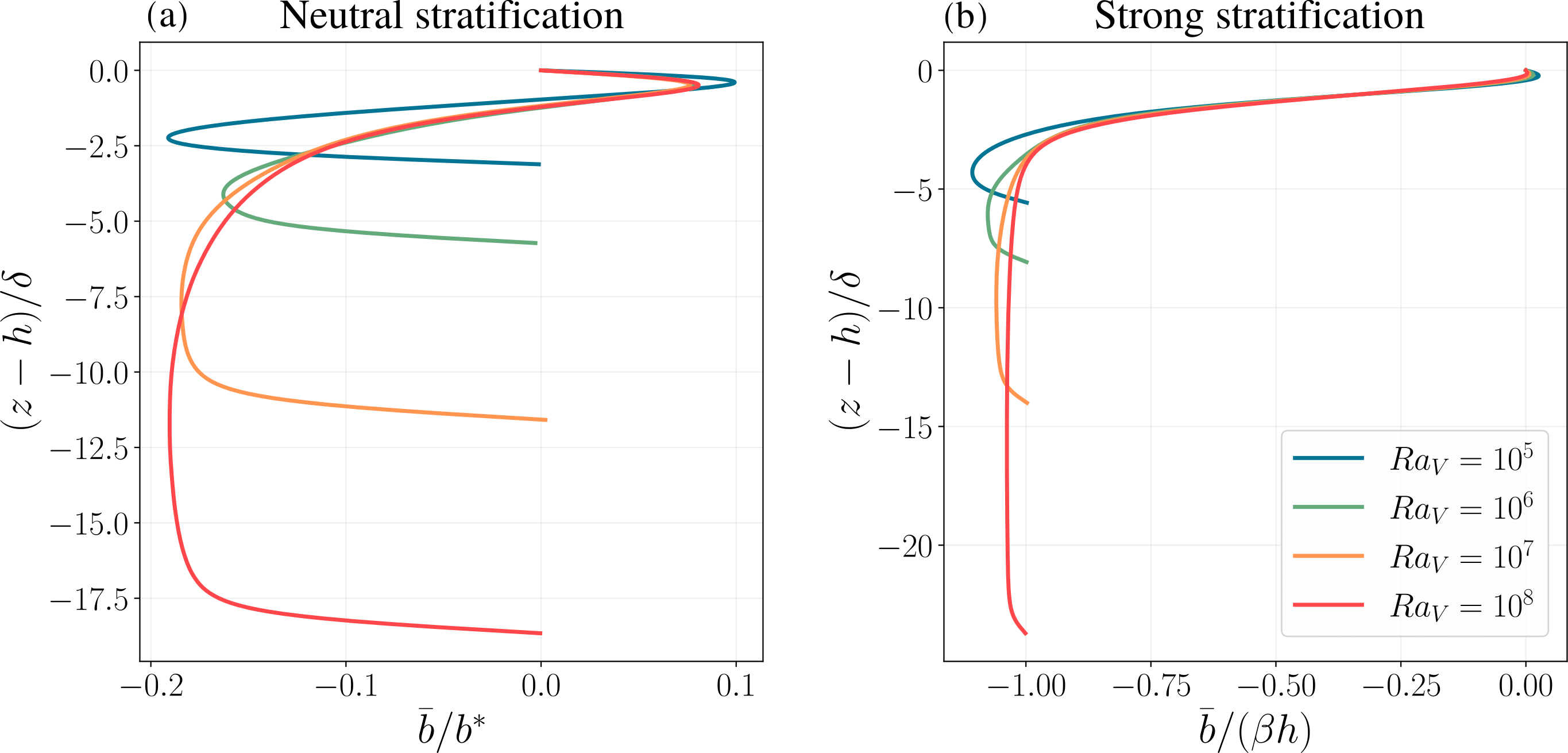}
    \caption{Vertical profile of the horizontal averaged buoyancy for several $\RaV$, normalized by $\bstar$ in the neutral stratification regime in $(a)$ and by $\beta h$ in the strong stratification regime in $(b)$. The boundary layer thickness $\delta$ is computed from the vertical profile of $\bar b_z(z)$  by determining the depth at which $95\%$ of the maximum value of $\bar b_z$ is reached.
    The parameters are $\Gamma = 8$ and $Pr=1$.}
  \label{fig4}
\end{figure}

 In the strong stratification regime RBC is largely suppressed: see  \autoref{fig3}(b) and \autoref{fig4}(b) in which $\la b_z\ra \approx + \beta$,  i.e.~this is the onset of the strongly restratified regime.
The horizontal bottom flow induced by the descending plumes at the top surface is sustained rather than intermittent. Some residual RBC effects remain, e.g.~in \autoref{fig3}(b) weak upward plumes generated by the bottom flux are  visible. These small RBC bottom plumes perturb the horizontal flow and enhance mixing near the bottom boundary compared to pure HC.
The structure of this flow is primarily determined by HC but retains some influence of RBC, particularly near the heated bottom and the lateral edges.

Vertical profiles of the horizontally averaged buoyancy  $\bar b(z)$ for several values of $\RaV$ are shown in \autoref{fig4}(b).
The buoyancy is normalized by the bottom forcing scale $\beta h$.
As in the neutral case, the vertical coordinate is rescaled by the thickness of the top boundary layer $\delta$, defined as the depth at which $\bar b_z$ reaches $95\%$ of its maximum value.

In this strongly restratified  regime, with $\langle b_z \rangle \approx +\beta$, the horizontally averaged buoyancy at the bottom is $\bar b(0) = -\beta h$.
When normalized by $\beta h$, $\bar b(z)$ profiles collapse near the top boundary -- see \autoref{fig4}(b) -- indicating that the buoyancy variation across the top boundary layer is now controlled by $\beta$ and confirming that this normalization of $\bar b$ captures the structure of the top boundary layer at the onset of the strongly restratified state.

\subsection{Transition Rayleigh numbers}
\begin{figure}
  \centering
  \includegraphics[width=1.0\textwidth]{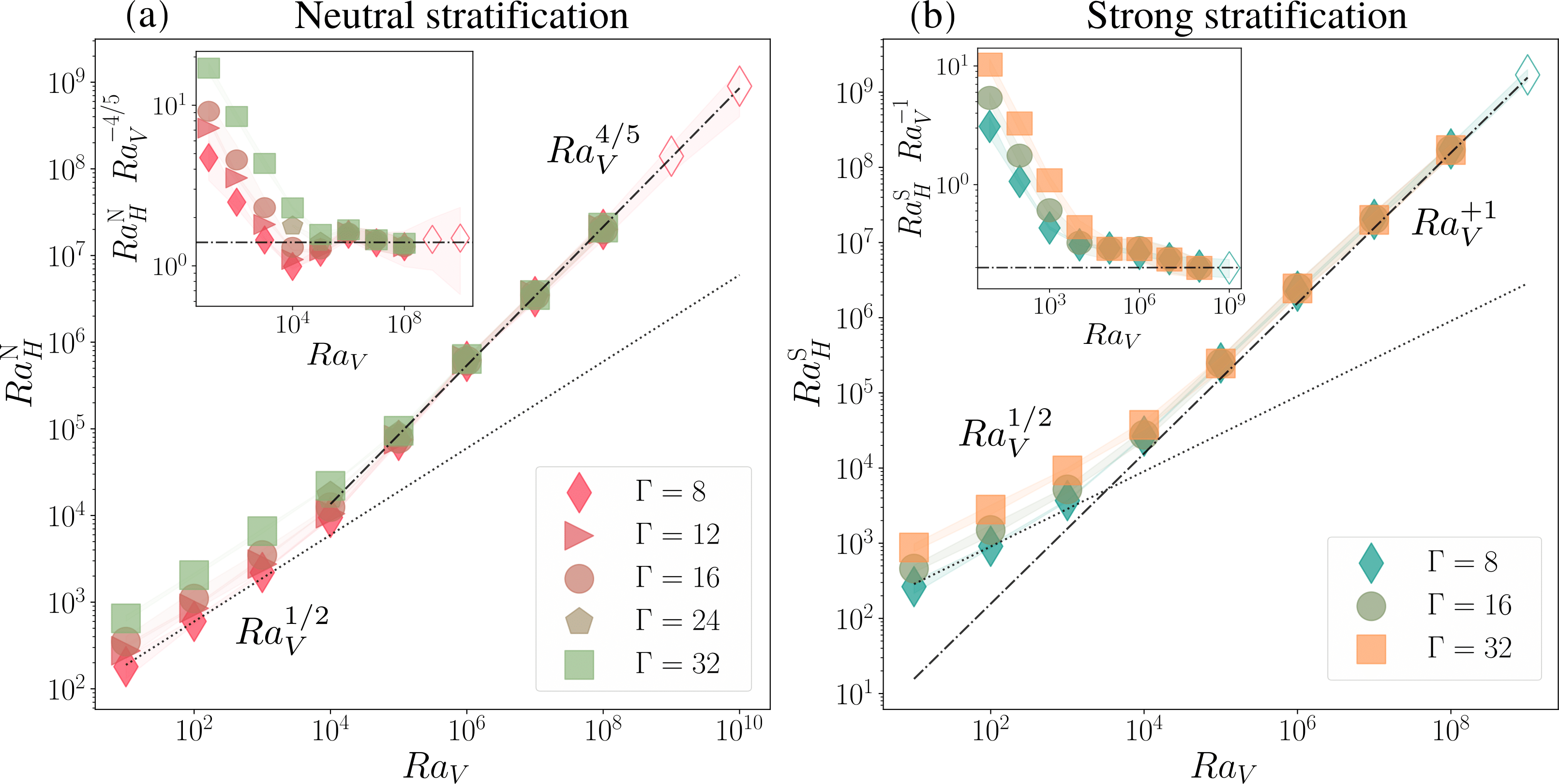}
    \caption{ $(a)$ Neutral horizontal Rayleigh number $\RaHN$ (defined as the value of $\RaH$ for which $\langle b_z \rangle = 0$ at a given $\RaV$) plotted as a function of $\RaV$ for several aspect ratios. The inset shows the same data compensated by the scaling law \eqref{rahincp}, For the two last $\RaV$ values ($10^9$, $10^{10}$), data are shown only for $\Gamma=8$. $(b)$ Strong horizontal Rayleigh number $\RaHstrg$ (defined as the value of $\RaH$ for which $\langle b_z \rangle = \beta$ at a given $\RaV$) plotted as a function of $\RaV$ for several aspect ratios. The inset shows the same data compensated by the scaling law \eqref{rahstrong}. For the last $\RaV$ value ($10^9$), data are shown only for $\Gamma=8$. Empty/full symbols indicate respectively filtered/DNS simulations.}
  \label{fig5}
\end{figure}

Consider a flow forced with a fixed value of $\RaV$. If $\RaH=0$ (pure RBC), then from \eqref{buoyDiss} $\la b_z\ra<0$.
As $\RaH$ is increased from zero, the flow initially remains unchanged and continues to be dominated by RBC.  
With further increases in $\RaH$ the system undergoes a sharp  transition: the convection pattern alternates between RBC cells and a large horizontal circulation driven by intermittent descending plumes from the upper boundary.  This transition corresponds to the regime change identified and discussed by CNF.
Further increasing $\RaH$ raises $\langle b_z\rangle$, and at some HC Rayleigh number, denoted $\RaHN$,  the flow is neutrally stratified with $\la b_z\ra =0$.
With further increase in $\RaH$  the flow becomes strongly restratified at $\RaHstrg > \RaHN$. To precisely characterize these transitions we define the horizontal Rayleigh numbers corresponding to the neutral and strong stratification states respectively as
\begin{align}
&\RaHN(\RaV,\Gamma,\Pr): \quad \RaH \ \text{at which} \ \langle b_z \rangle = 0, \\
&\RaHstrg(\RaV,\Gamma,\Pr): \quad \RaH \ \text{at which} \ \langle b_z \rangle = \beta.
\end{align}
\autoref{fig5} shows the functions $\RaHN$ and $\RaHstrg$ in the $(\RaH,\RaV)$ parameter space for $Pr=1$; we consider several aspect ratios as indicated.
Each point in \autoref{fig5}  is determined using a bisection search: to determine $\RaHN$ we fix  $\RaV$, $\Gamma$ and $\Pr$. We then obtain a sequence of solutions in which the target $\RaHN$ is bracketed and repeatedly halved until we located the value of $\RaH$ for which $|\la b_z\ra /\beta|\leq0.05$.
The shaded regions in \autoref{fig5} indicate the  uncertainty.

For low Rayleigh numbers ($\RaH, \RaV \lesssim 10^4$), both neutral stratification and the onset of the strong stratification regime follow a power-law scaling  

\beq (\RaHN,\RaHstrg)\sim\RaV^{1/2} \Gamma
\label{lowRa1}
\eeq
with no dependence on $\Pr$. Thus a main  conclusion is that the neutral curve $\RaH=\RaHN(\RaV,\Gamma,\Pr)$ passes through the origin of the $(\RaV,\RaH)$ parameter plane. Further detail and an analytic explanation of \eqref{lowRa1} are in \autoref{appLowRa}. 

At higher Rayleigh numbers ($\RaH, \RaV \gtrsim 10^4$), the system enters an aspect-ratio-independent regime, with 
\beq
\RaHN(\RaV,\Gamma,1) \sim 8.54~\RaV^{4/5}\per
\label{neutscal}
\eeq
The onset of the strong stratification regime is at
\beq
\RaHstrg(\RaV,\Gamma,1) \sim 1.70~\RaV \per
\label{strongscal}
\eeq
The scaling above is consistent with the conclusion following \eqref{strong1} that strong restratification requires $\RaH>\RaV$.

The regimes in \eqref{neutscal} and \eqref{strongscal}  are independent of aspect ratio $\Gamma$. This underscores the relevance of the layer thickness $h$ as the characteristic length scale of the system and motivates the definition of $\RaH$ in \eqref{RaZ} --  using $\RaHtrad$ in \eqref{RaTrad} introduces an artificial dependence on~$\Gamma$.
Figure~\ref{fig5}(a) and (b) show our estimates of $\RaHN$ and $\RaHstrg$ at $\Pr=1$ for different values of the aspect ratio $\Gamma$.

\section{Scaling analysis}\label{sec:scalings}
In this section, we analyze the system using scaling arguments, focusing on regimes where either RBC or HC dominates, as well as the neutral state with $\la b_z \ra=0$ and the onset  of strong stratification at $\la b_z \ra=+\beta$.
The horizontally-averaged buoyancy equation \eqref{PY1} shows that the buoyancy flux entering the domain through the bottom boundary layer must be conserved across any horizontal plane.
Consequently, buoyancy must be extracted through a boundary layer at the top surface, leading to  $\bar{b}_z(h) = -\beta$.
Due to the imposed buoyancy profile at the top boundary, we expect this upper region to be the primary site where the competition between RBC and HC manifests itself, thereby governing the global dynamics of the system. We therefore focus on the structure and behavior of the top boundary layer.

Most of our numerical solutions use $\Pr=1$ and  we do not distinguish between viscous and buoyancy boundary layers.
We assume that the characteristic vertical length scale is given by the thickness of the top boundary layer, denoted by $\delta$, such that $\partial/\partial_z \sim 1/\delta$.
The characteristic horizontal length scale is taken to be $\ell$, which will be specified for each regime considered, implying $\partial/\partial_x \sim 1/\ell$.

We expect that a significant portion of both viscous and buoyancy dissipation is concentrated within the boundary layers (BLs), which occupy a fraction $\delta/h$ of the domain. Accordingly, we estimate the dissipation rates as
\begin{equation}
    \varepsilon = \nu \langle |\nabla \mathbf{u}|^2 \rangle \sim \nu \frac{U^2}{\delta^2} \left( \frac{\delta}{h} \right), \quad \chi = \kappa \langle |\nabla b|^2 \rangle \sim \kappa \frac{B^2}{\delta^2} \left( \frac{\delta}{h} \right),
    \label{dissscal}
\end{equation}
where $U$ is the characteristic horizontal velocity and $B$ is the buoyancy difference across the top boundary layer, with the factor $(\delta/h)$ coming from the ratio of the boundary layer thickness to the total depth.

From mass conservation \eqref{divu}, we obtain the relation $U/\ell \sim W/\delta$, where $W$ is the characteristic vertical velocity within the top BL.
Balancing advection with diffusion in the buoyancy equation \eqref{buoy} and this relation yields a scaling for the horizontal velocity
\begin{equation}
    U \sim \frac{\kappa \ell}{\delta^2}.
    \label{Uscal}
\end{equation}

\subsection{Scalings in the asymptotic regimes}
We now derive scaling laws for the top boundary layer thickness and mean buoyancy in two asymptotic regimes: when RBC dominates HC ($\RaV \gg \RaH$), and when HC dominates RBC ($\RaH \gg \RaV$).
\subsubsection{RBC-dominated regime}
\begin{figure}
  \centering
  \includegraphics[width=1.0\textwidth]{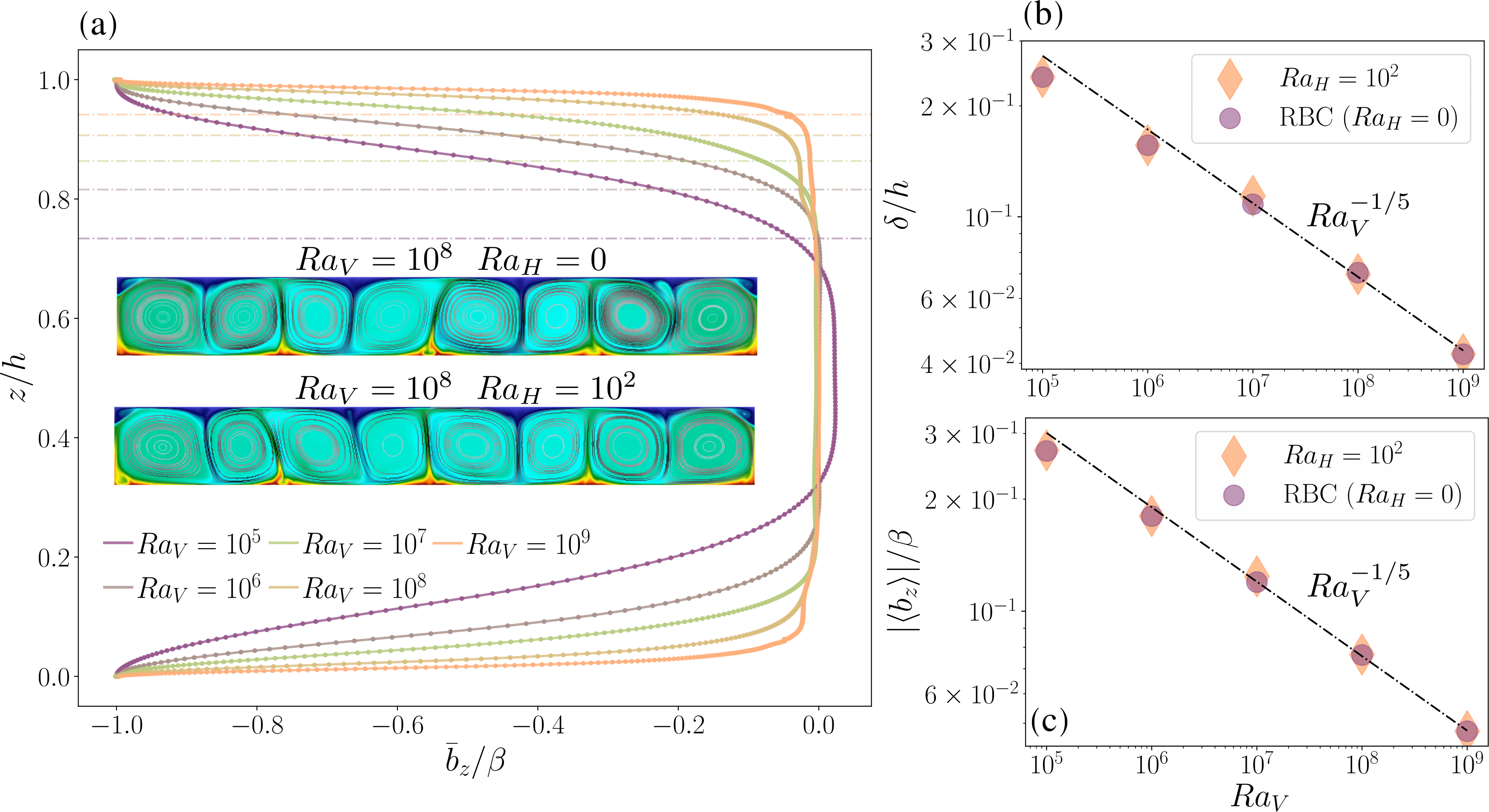}
    \caption{Vertical profiles of the mean vertical buoyancy gradient, $\bar{b}_z$, for varying $\RaV$ with fixed $\RaH = 10^2$ in panel $(a)$. Dash-dot lines indicate the estimated thickness of the top boundary layer. Insets show snapshots of the buoyancy field: in $(a)$, for pure RBC (top, $\RaV = 10^8$) and a mixed case (bottom, $\RaV = 10^8$, $\RaH = 10^2$). Panel $(b)$ shows the boundary layer thickness $\delta$ as a function of $\RaV$ for both pure RBC and mixed case. Panel $(c)$ shows the norm of the mean vertical buoyancy gradient, $|\langle b_z \rangle|$, and its dependence on $\RaV$ for both pure RBC and mixed cases.}
  \label{fig6}
\end{figure}
When the flux Rayleigh number greatly exceeds the horizontal Rayleigh number ($\RaV \gg \RaH$), the dynamics are controlled by RBC.  
The horizontal length scale is set by the depth of the layer, $\ell \sim h$, and the imposed bottom buoyancy flux dominates the surface forcing.  
In this limit the power integrals \eqref{veps2}--\eqref{buoyDiss} reduce to
\begin{equation}
    \varepsilon \approx \kappa \beta \quad\mbox{and} \quad \chi \approx -\kappa \beta \langle b_z \rangle.
    \label{dissRBC}
\end{equation}
Assuming $\la b_z \ra \sim B/h$, i.e.~$(B /\delta)$ times the global average factor $(\delta/h)$ and substituting the dissipation scalings \eqref{dissscal} into \eqref{dissRBC} yields
\begin{equation}
    \delta \sim h \RaV^{-1/5}, \quad B \sim \beta \delta \sim \beta h \RaV^{-1/5}.
    \label{RBscaling}
\end{equation}
These scalings are the fixed-flux equivalent of the regime-I scalings of \cite{GL2000} for fixed-temperature boundaries, namely $\delta\sim h Ra_{\Delta b}^{-1/4}$. where $Ra_{\Delta b}$ is the classical RBC parameter based on the buoyancy difference $\Delta b$ between the top and bottom plate.
Adapting $Ra_{\Delta b}^{-1/4}$ to an an imposed bottom flux using $\Delta b \approx \beta\delta$ gives $Ra_{\Delta b} = (\delta/h)\RaV$ \citep{otero2002,johnston2009}, which then recovers \eqref{RBscaling} for the boundary-layer thickness.

The predicted trends are confirmed in \autoref{fig6}(b) and (c).  
Profiles of $\bar b_z$ (panel a) show the definition of the top boundary-layer thickness $\delta$, taken as the distance from the top where $\bar b_z$ falls to $95\%$ of its maximum. 
In these panels, we present data from pure RBC simulations at $\RaV = 10^8$ and from a mixed-forcing case with $\RaV = 10^8$ and $\RaH = 10^2$.
For the mixed case with $\RaV=10^8$ and $\RaH=10^2$, the data for both $\delta$ and $|\langle b_z \rangle|$ collapse onto those from pure RBC, indicating that the dynamics remain governed by RBC, with the horizontal forcing being too weak to produce any significant modification (see inset in \autoref{fig6}(a)).

\subsubsection{HC-dominated regime}
\begin{figure}
  \centering
  \includegraphics[width=1.0\textwidth]{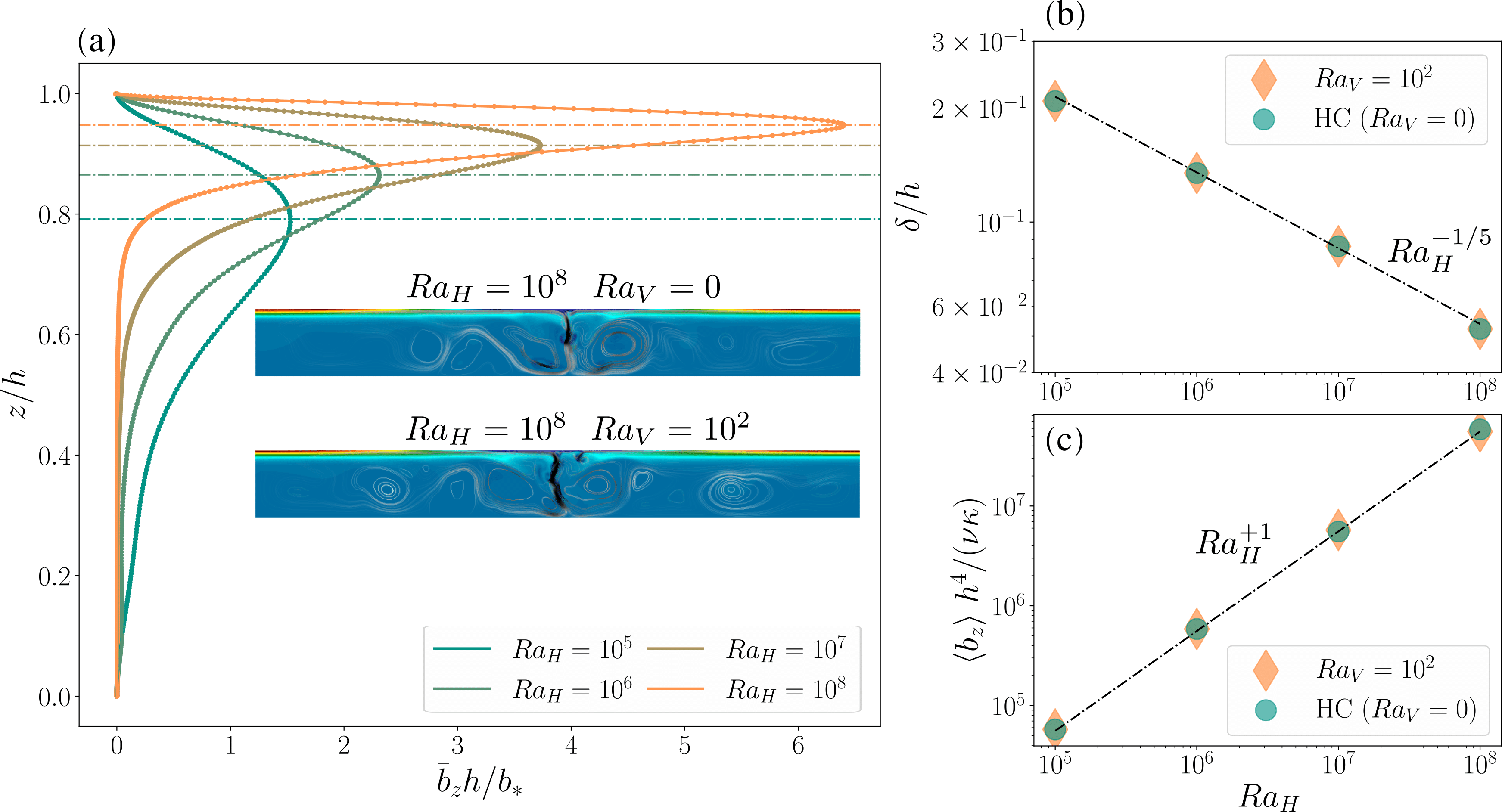}
    \caption{Vertical profiles of the mean vertical buoyancy gradient, $\bar{b}_z$, for varying $\RaH$ with fixed $\RaV = 10^2$ in panel $(a)$. Dash-dot lines indicate the estimated thickness of the top boundary layer. Insets show snapshots of the buoyancy field for pure HC (top, $\RaH = 10^8$) and a mixed regime (bottom, $\RaH = 10^8$, $\RaV = 10^2$). Panel $(b)$ shows the boundary layer thickness $\delta$ as a function of $\RaH$ for both pure HC and mixed regime. Panel $(c)$ shows the mean vertical buoyancy gradient $\langle b_z \rangle$, and its dependence in $\RaH$ for both pure HC and mixed regimes.}
  \label{fig7}
\end{figure}
When the horizontal Rayleigh number is much larger than the flux Rayleigh number ($\RaH \gg \RaV$), the flow is governed by HC.  
The relevant horizontal scale is that of the imposed surface buoyancy profile, $\ell \sim \ellx$, and the surface forcing overwhelms the bottom flux.  
In this limit the power integrals reduce to
\begin{equation}
    \varepsilon \approx \kappa \langle b_z \rangle,
    \qquad
    \chi \approx \frac{\kappa}{h}\,\overline{b_z(h)\,\bs(x)} .
    \label{dissHC}
\end{equation}
Using the dissipation estimates \eqref{dissscal} gives
\begin{equation}
    \delta \sim \ellx \left( \RaH \Gamma^3 \right)^{-1/5},
    \qquad
    B \sim \bstar \sim \frac{\nu \kappa}{h^3}\,\RaH .
    \label{HCscal}
\end{equation}
The explicit dependence on aspect ratio $\Gamma$ reflects the fact that the HC Rayleigh number is naturally defined with the horizontal scale, $Ra_{\ellx} = \RaH \Gamma^3$.  
These scalings coincide with the classical HC results of \cite{rossby1965}; see also the review of \cite{ANRHC2008}.

The agreement with numerical solutions is shown in \autoref{fig7}(b) and (c).  
In these panels, we show data from pure HC simulations at $\RaH = 10^8$ and from a mixed-forcing case with $\RaH = 10^8$ and $\RaV = 10^2$.
For the mixed case, the measured $\delta$ and $|\langle b_z \rangle|$ collapse onto those from pure HC, indicating that the weak bottom forcing produces no measurable departure from HC behaviour (see inset of \autoref{fig7}(a)).

\subsection{Neutral stratification state}
Having examined the asymptotic regimes dominated by either RBC or HC, we now turn to the intermediate case corresponding to neutral stratification, defined by $\langle b_z \rangle = 0$.
We adopt the same scaling approach as in the asymptotic regimes.
In this transitional state, where neither forcing clearly dominates, the characteristic horizontal length scale is expected to scale with the domain height, i.e.\ $\ell \sim h$.
Importantly, with the condition $\langle b_z \rangle = 0$, the power integrals \eqref{veps2} and \eqref{buoyDiss} simplify to two-term expressions:
\begin{equation}
    \varepsilon = \kappa \beta \quad \mbox{and} \quad \chi = \frac{\kappa}{h} \, \overline{b_z(h) \, \bs(x)}.
    \label{dissNeutral}
\end{equation}
Reflecting the mixed nature of the neutral stratification state, $\varepsilon$ has the same expression as in the RBC-dominated regime \eqref{dissRBC}, while $\chi$ has the HC- form in \eqref{dissHC}.
Substituting the dissipation scalings from \eqref{dissscal} into \eqref{dissNeutral} yields the following relations for the top boundary layer thickness and buoyancy jump:
\begin{equation}
    \delta \sim h \RaV^{-1/5}, \quad B \sim \bstar \sim \frac{\nu \kappa}{h^3} \RaH.
    \label{Neutralscal}
\end{equation}
Thus, the BL thickness follows the RBC scaling, while the buoyancy difference across the BL follows the HC scaling.
Moreover the scaling for the buoyancy jump is consistent with \autoref{fig4}(a).
However, since the net buoyancy flux must exit through the top boundary layer, the horizontally averaged buoyancy equation \eqref{PY1} still imposes the condition $\bar{b}_z \approx -\beta$. Using this, along with the scaling $\bar{b}_z \sim \bstar/\delta$, we obtain:
\begin{equation}
    \bstar \sim \beta \delta.
    \label{diff2}
\end{equation}
Multiplying both sides of \eqref{diff2} by $h^4 / (\nu \kappa)$ and substituting the expression for $\delta$ from \eqref{Neutralscal} leads to the following relation:
\begin{equation}
    \RaHN \sim \RaV^{4/5}.
    \label{rahincp}
\end{equation}
This scaling is in excellent agreement with the transition identified in \autoref{fig5}(a), thereby confirming the theoretical prediction.


\subsection{The onset of strong stratification}
We now consider the onset of the strong stratification, defined by the condition that $\langle b_z \rangle = +\beta$, i.e.~the global stratification is equal in magnitude, but opposite in sign, to $\bar b_z$ at the top and bottom of the layer. 
In this regime, HC is expected to dominate the dynamics, although not fully, meaning the characteristic horizontal length scale remains set by the domain height, $\ell \sim h$, rather than the imposed surface scale $\ellx$ (see \autoref{fig3}).

With $\langle b_z \rangle = +\beta$ the power integrals \eqref{veps2} and \eqref{buoyDiss}  simplify to:
\begin{equation}
    \varepsilon = 2\kappa \beta, \quad \text{and} \quad \chi = \frac{\kappa}{h} \, \overline{b_z(h) \, \bs(x)} - \kappa\beta^2.
    \label{dissStrong}
\end{equation}
As in the neutral regime, $\varepsilon$ follows the RBC scaling \eqref{dissRBC}, while the buoyancy variance dissipation $\chi$ includes contributions from both RBC and HC.
Substituting the dissipation scalings from \eqref{dissscal} into \eqref{dissStrong} yields the following expressions for the top boundary layer thickness and buoyancy jump:
\begin{equation}
    \delta \sim h \RaV^{-1/5}, \quad B \sim \bstar - \frac{\beta^2 \delta h}{B}.
    \label{Strongscal}
\end{equation}
The boundary layer thickness continues to follow the RBC scaling in \eqref{Neutralscal}, while the buoyancy jump $B$ is modified by a combination of HC and RBC effects.
Because $\langle b_z\rangle=\beta$ and  $\langle b_z\rangle \sim B/h$, it follows that $B \sim \beta h$, in agreement with the scaling shown in \autoref{fig4}(b).
Substituting $B \sim \beta h$ into the expression for $B$ in \eqref{Strongscal} leads to
\begin{equation}
    \bstar \sim \beta h \left(1 + \frac{\delta}{h} \right).
    \label{diff3}
\end{equation}
Multiplying both sides of \eqref{diff3} by $h^3 / (\nu \kappa)$ and using the RBC scaling for $\delta$ from \eqref{Neutralscal}, we obtain the following scaling:
\begin{equation}
    \RaHstrg \sim \RaV \left(1 + \RaV^{-1/5} \right).
    \label{rahstrong}
\end{equation}
The relation above reveals two distinct contributions to $\RaHstrg$: a dominant term proportional to $\RaV$ and a correction scaling with $\RaV^{4/5}$. 
In the asymptotic limit $\RaV \gg 1$, the leading-order behavior is  $\RaHstrg \sim \RaV$. This scaling is in  agreement with the transition identified in \autoref{fig5}(b).

The scaling arguments above apply to the regime in which the flow remains boundary-layer dominated (regime~I of \cite{GL2000}). At higher Rayleigh numbers, transitions to different  regimes of convection and associated changes in the Nusselt–Rayleigh scaling are expected. These will likely modify the $\RaHN$ and $\RaHstrg$ scalings above.

\section{Discussion \label{discussion}}

\subsection{Sensitivity to Prandtl number}
\begin{figure}
  \centering
\includegraphics[width=0.90\textwidth]{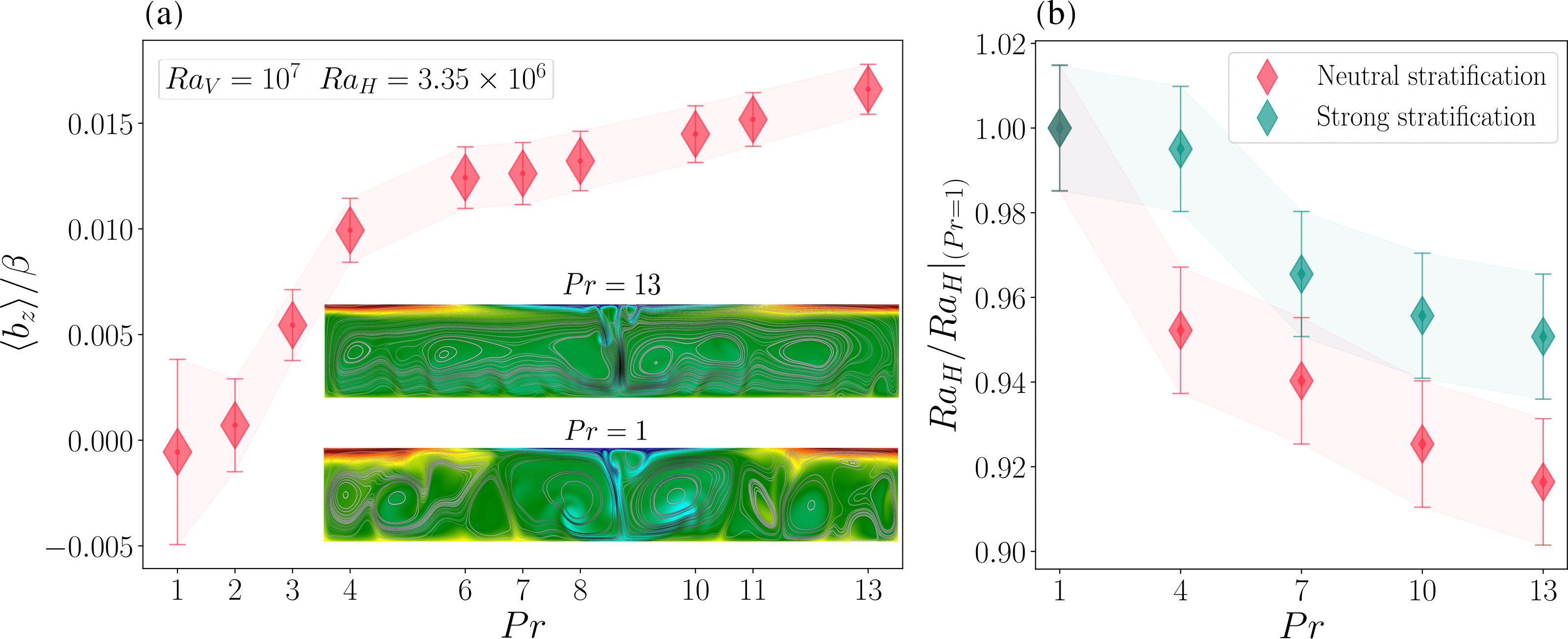}
    \caption{(a) Volume-averaged vertical buoyancy gradient, normalized by the imposed bottom flux $\beta$, as a function of the Prandtl number at the neutral state ($\RaH = 3.35\times10^{6}$).  
Insets show snapshots of the buoyancy field for $Pr = 13$ (top) and $Pr = 1$ (bottom).  
Error bars and the shaded region indicate one standard deviation about the temporal mean.  
(b) Neutral and strong horizontal Rayleigh numbers, each normalized by their value at $Pr = 1$, plotted versus the Prandtl number. Error bars and the shaded region indicate the uncertainty.  All data correspond to $\RaV = 10^{7}$ and $\Gamma = 8$.}
  \label{fig8}
\end{figure}

Computations in the previous sections used $Pr=1$. But the Prandtl number of water in icy-moon conditions is in the range  10 to 13.
\autoref{fig8}(a) shows the variation of $\langle b_z \rangle / \beta$ with increases in $Pr$.
Starting from the neutral stratification regime at $\RaV=10^7$ ($\RaH=\RaHN=3.35\times10^6$) and $(\Gamma,\Pr)=(8,1)$  we  increase $Pr$ with both Rayleigh numbers fixed.
For each case, the system is evolved to a statistically steady state, and $\langle b_z \rangle$ is computed by averaging over at least one diffusive time.

Increasing $Pr$  enhances restratification: at   $Pr=13$, $\langle b_z \rangle / \beta = 0.015$. 
The insets in \autoref{fig8}(a) show that at $Pr=13$ the flow  exhibits features closer to the strong stratification regime e.g.~convection cells are further suppressed and bottom plumes are weakened. \autoref{fig8}(b) shows the dependence of $\RaHN$ and $\RaHstrg$ on $\Pr$. Increasing $Pr$ lowers both $\RaHN$ and $\RaHstrg$ i.e.~at larger $\Pr$ a weaker horizontal buoyancy forcing is required to attain $\RaHN$ and $\RaHstrg$ at fixed $\RaV$. But the effect is small: between $Pr=1$ and $Pr=13$, $\RaHN$ decreases by only about 10\% and $\RaHstrg$ by roughly 5\%.
Although this decrease in $\RaHN$ and $\RaHstrg$ is modest, we conclude that increasing $\Pr$ increases restratification by HC.

The dependence of the neutral transition on $Pr$ may vary outside the moderate range considered here ($1\leq Pr\leq 13$), where different transport regimes are expected in both RBC \citep{GL2001,lindborg2023scaling} and HC \citep{Shishkina2016,Passaggia_Scotti_2024,Passaggia_Cohen_Scotti_2024}. Within the range $1\leq Pr\leq 13$, we interpret the observed trend qualitatively in terms of boundary-layer structure: increasing $Pr$  thins the buoyancy boundary layer relative to the viscous boundary layer, reducing the buoyancy contrast across it and hence the horizontal forcing required to attain neutral stratification.


\subsection{Comparison with CNF}
CNF's configuration differs from ours in two respects. Their half-wavelength sinusoidal surface forcing produces a single HC cell, whereas our full wavelength sinusoidal forcing in \eqref{sbuoy17} results in two counter-rotating HC cells. 
Second, our HC plume does not interact with the no-slip sidewalls of the domain, whereas CNF's plume does. In pure HC this second difference produces a compact recirculating eddy associated with the HC plume; e.g.~see CNF figure 2(f). This recirculating eddy is obtained in our configuration by making the sign of $\bstar$ in \eqref{sbuoy17} negative, equivalently $\RaH<0$. Quantifying the effect, if any, of $\RaH<0$ on restratification is beyond the scope of this work.

\begin{figure}
\centering
\includegraphics[width=0.9\textwidth]{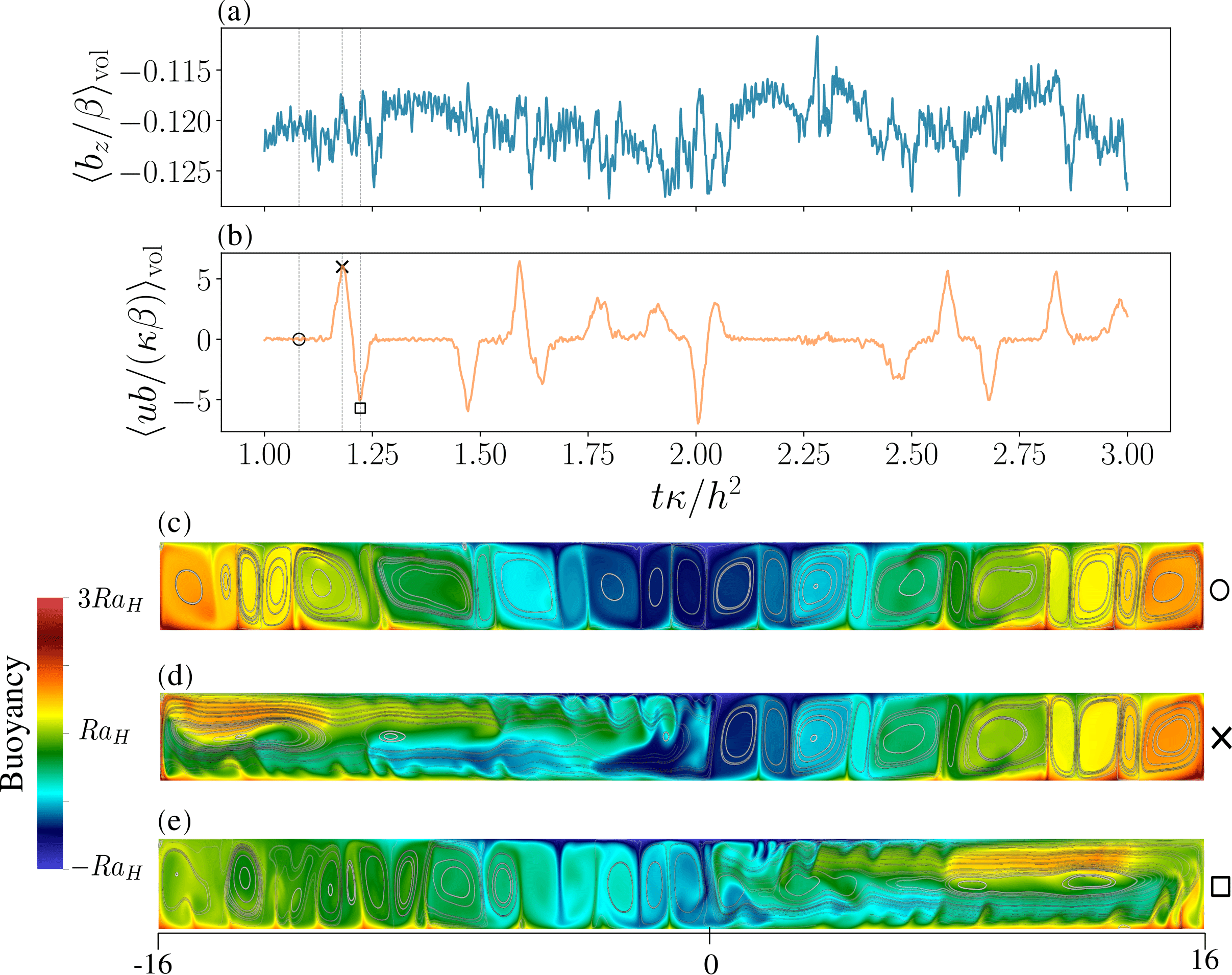}
\caption{Time evolution of volume‐averaged quantities (denoted  $\langle \,\cdot\, \rangle_{\mathrm{vol}}$):  
 $(a)$ $\langle u b \rangle_{\mathrm{vol}}/ (\kappa\beta) $, and $(b)$ $\langle b_z \rangle_{\mathrm{vol}}/ \beta $. 
Panels $(c)$, $(d)$ and $(e)$ show  snapshots  (corresponding to times indicated in panel (b))  of the buoyancy field. The bulk stratification is unstable with $\la b_z\ra/\beta=-0.121$.
Parameters are $(\RaH,\RaV) = (10^6,10^7)$ and $(\Gamma,\Pr) = (32,1)$.}
\label{fig9}
\end{figure}

Fixing $\RaV$, and varying an analog of  $\RaH$, CNF observed an interesting phenomenology involving the transient interruption of cellular RBC by episodic ``bursts'' of HC.
CNF identify bursting as characteristic of the transition between an RBC regime and an HC regime. This identification of the RBC-HC transition differs  from our $\la b_z\ra=0$ criterion for restratification. 

\autoref{fig9} shows a solution in our configuration exhibiting HC bursting. The main point is that the time series of the instantaneous volume average $b_z$ is always negative: see  \autoref{fig9}(a). In other words, as $\RaH$ is increased, with fixed $\RaV$ the system first enters the bursting regime, and then, at higher $\RaH$, restratifcation occurs.

The CNF bursts occupy the entire horizontal extent of the domain. But here, because of  full‐wavelength surface forcing, bursts of HC occur independently in either half of the domain: see \autoref{fig9} (c) (d) and (e).  
During these bursts of HC, the flow carries a significant horizontal buoyancy flux: see \autoref{fig9}(b). The volume‐averaged of $b_z$ in \autoref{fig9}(a) seems unaffected by this episodic horizontal buoyancy flux. 

Louis‐Alexandre Couston (personal communication)  notes that in the single-cell configuration of CNF, HC bursts also occur with $\langle b_z \rangle < 0$. Summary: bursting is a first indication that HC is challenging  RBC, but further increase in $\RaH$ is required for restratification.

\begin{figure}
\centering
\includegraphics[width=0.9\textwidth]{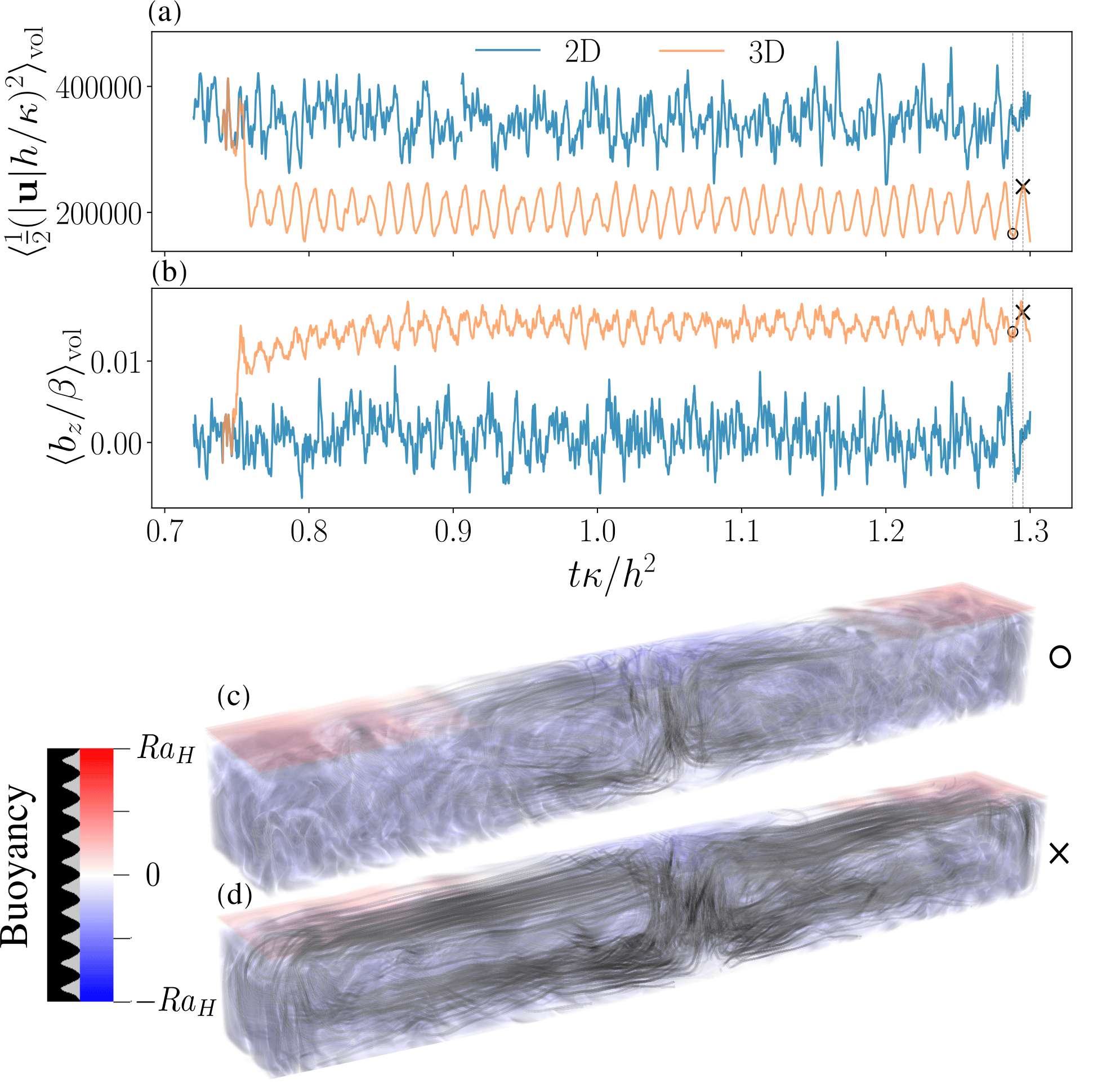}
\caption{Time evolution of volume‐averaged quantities: $(a)$ kinetic energy and $(b)$ $\la b_z / \beta\ra_{\mathrm{vol}}$.
Results from a 2D simulation are shown in blue, and the continuation of this case in 3D in orange.
Panels $(c)$ and $(d)$ show snapshots of the buoyancy field at the times indicated in panel $(a)$, with streamlines overlaid (colour scale and opacity shown in the colour bar).
The bulk stratification is stable with $\la b_z\ra/\beta=1.45 \times 10^{-2}$.
Parameters are $(\RaH,\RaV) = (0.2,1)\times 10^8$ and $(\Gamma,\Pr) = (8,1)$.}
\label{fig10}
\end{figure}

\begin{figure}
\centering
\includegraphics[width=0.9\textwidth]{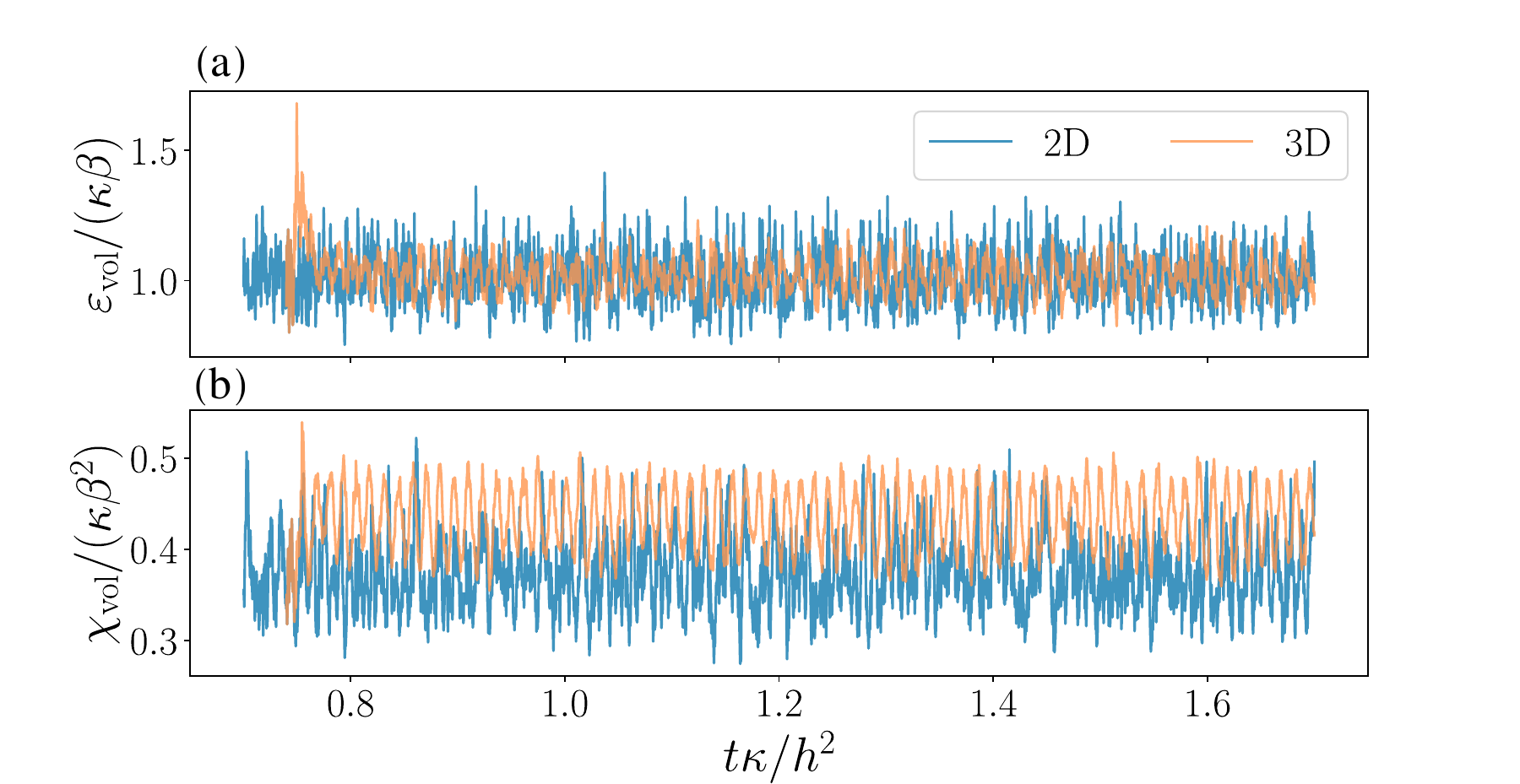}
\caption{Time evolution of volume‐averaged quantities: $(a)$ $\varepsilon_{\mathrm{vol}}/(\kappa \beta)$ and $(a)$ $\chi_{\mathrm{vol}}/(\kappa \beta^2)$
where $\varepsilon_{\mathrm{vol}}= \nu \la |\bm{\nabla u}|^2 \ra_{\mathrm{vol}}$ and 
$\chi_{\mathrm{vol}}= \kappa \la |\nabla b|^2 \ra_{\mathrm{vol}}$.
Results from an effectively 2D simulation ($\ell_y/h=1/8$) are shown in blue, and the continuation of this case in 3D ($\ell_y/h=1$) in orange.
Parameters are $(\RaH,\RaV) = (0.2,1)\times 10^8$ and $(\Gamma,\Pr) = (8,1)$.}
\label{fig10_2}
\end{figure}

\subsection{Three-dimensionality: through thick and thin}
Convection can differ significantly between two and three dimensions. In particular, the HC sweeping mechanism associated with the onset of the neutral state may be modified in 3D. Furthermore, while the inverse cascade of 2D turbulence promotes the formation of large-scale coherent structures, 3D flows exhibit a direct cascade of kinetic energy towards smaller scales and enhanced viscous dissipation. 

To assess these effects, we undertook additional 3D simulations with periodic boundary conditions in the transverse $y$-direction. 2D solutions correspond to  $\ell_y=0$. With $\ell_y=h/16$ 
(a `thin' 3D domain) the  numerical solution evolves to become  independent of $y$ and remains so; i.e. this thin 3D solution is statistically identical to the corresponding 2D solution with $\ell_y=0$. Once the thin 3D solution reaches a statistically steady state we use a single-time snapshot as a 2D  initial condition for a `thick' 3D numerical solution with $\ell_y= h$. The thick 3D numerical solution is explosively unstable to transverse  3D disturbances and evolves into a complex eddying 3D flow. (We did not attempt to determined the critical value of $\ell_y$ that triggers `three-dimensionalization'.)   

Figures \ref{fig10} and \ref{fig10_2}  compare the evolution of the thick 3D solution ($\ell_y=h$) to a thin ($\ell_y=h/16$) neutral ($\la b_z/\beta\ra \approx 8.9\times 10^{-4}$) solution.  Activation of the third dimension decreases the kinetic energy by about 40\% (figure \ref{fig10}(a)). The mean stratification increases slightly, with $\langle b_z\rangle$ rising to about $0.015 \beta$ (figure \ref{fig10}(b)). The viscous power integral
\beq
\varepsilon = \kappa \beta + \kappa \la b_z\ra\com
\eeq
indicates that with fixed $\beta$ an increase in $\la b_z\ra$ must be accompanied by an increase in viscous dissipation $\varepsilon$.
This increase in $\varepsilon$ is masked, however, by large fluctuations in the $\varepsilon_{\mathrm{vol}}(t)$ time series: see  \autoref{fig10_2}(a).
The 3D flow also has  approximately 15\% larger buoyancy variance dissipation $\chi$ than that of the 2D comparison flow (\autoref{fig10_2}(b)). From the buoyancy variance power integral  \eqref{buoyDiss}, for fixed $\beta$ this increase in $\chi$ requires an increase of the surface correlation term $\overline{b_z(h)b^s(x)}/h$ sufficient to compensate for the larger mean stratification $\langle b_z\rangle$.

The time series from the 3D  solution  in \autoref{fig10}  has striking oscillatory behaviour. This noisy periodicity is particularly clear in the kinetic energy time series  in  \autoref{fig10}(a).
These oscillations are associated with the main descending central plume and its associated large-scale cellular return flow  that sweeps small bottom rising plumes towards the lateral boundaries (see \autoref{fig10}(c) and (d)). In the high kinetic energy state, denoted by $\times$ in  \autoref{fig10}, the domain scale HC cells on either side of the central  plume occupy the whole domain. In the low kinetic energy state, denoted by $\circ$, the HC cells are restricted to the centre of the domain. Outside the central region the large-scale flow is weak and disorganized: see \autoref{fig10}(c). Thus the noisy periodicity in kinetic energy corresponds to a expansion and contraction of the HC cells.

Despite some  quantitative differences, most notably lower kinetic energy,  the  3D flow with $\ell_y/h=1$ is similar to that of the 2D flow with $\ell_y/h =0$ (and the effectively 2D flow with $\ell_y/h=1/16$).  
The modest increase in $\langle b_z\rangle$ indicates that the small additional dissipation $\varepsilon$ introduced by three-dimensional motions remains secondary compared with the dominant boundary-layer dissipation. The restratification mechanism remains unchanged.

Three-dimensional simulations with $\RaV = 10^6$ and $10^7$ showed the same qualitative behavior as for $\RaV=10^8$ discussed above.

\section{Conclusion}
This study identifies the dynamical conditions under which bottom‐heated fluids become restratified and shows that HC plays a central role in setting the global buoyancy structure.  
By linking the onset of restratification to the competition between vertical and horizontal buoyancy forcing, we provide a framework for predicting how global stratification develops and adjusts.
Future work should extend this framework to include the effects of rotation, salinity, spatially inhomogeneous bottom heating and spherical geometry, which are essential ingredients for assessing restratification processes in subglacial  oceans and lakes.

\backsection[Acknowledgements] {The authors
thank Louis-Alexandre Couston for discussion of this problem and 
the referees for their constructive comments, which improved the manuscript.}

\backsection[Funding]{This research was supported by the Simons Foundation as part of the project \textit{Fundamental Fluid Processes in Climate, Stellar, and Planetary Modeling}.}

\backsection[Declaration of interests]{The authors report no conflict of interest.}


\backsection[Author ORCID]{Florian Rein, https://orcid.org/0009-0006-2060-7362. Stefan Llewellyn Smith, https://orcid.org/0000-0002-1419-6505.  William~Roy Young,  https://orcid.org/0000-0002-1842-3197.}

\appendix
\section{The geothermal  flux Rayleigh number \label{fluxRa}} 
As a typical geothermal heat flux, $Q^{\text{gt}}$, we use 50 mW m$^{-2}$ (milliwatts per square meter), characteristic of abyssal-plane heat flux,  but less than   the global mean 86.4 mW\,m$^{-2}$ estimated by \cite{emile2009}.  If $Q^{\text{gt}}$ is  transmitted vertically  only by molecular diffusion of heat in water, with diffusivity $\kappa$, then the temperature gradient is
\beq
T_z^{\text{diff}} = \frac{Q^{\text{gt}}}{\kappa \rho c_p} \per
\label{Qgt}
\eeq
In the denominator above $\kappa \rho c_p$ is the thermal conductivity, denoted $k$ in CNF.
With $\rho c_p \approx 4\times 10^6$J m$^{-3}$ K$^{-1}$ and $\kappa \approx 10^{-7}$m$^2$ s$^{-1}$, the vertical temperature gradient in \eqref{Qgt}  is $0.125$K m$^{-1}$. With an ocean depth $h\approx5000$ m (characteristic of an abyssal plane) the implied vertical temperature difference between top and bottom is
\beq
\Delta T_{\mathrm{V}}^{\text{diff}} = hT_z^{\text{diff}} = 625K.
\eeq
Although the 50 milliwatts per square meter geothermal  flux  is smaller than ocean surface heat fluxes by a  factor of $10^3$ or $10^4$, $\Delta T_{\mathrm{V}}^{\text{diff}} = 625$K indicates that unopposed geothermal forcing is sufficient to produce strongly supercritical RBC.

Using $\Delta T_{\mathrm{V}}^{\text{diff}}$ motivates the definition of the flux Rayleigh number as
\beq
\RaV = \frac{g \alpha \Delta T_{\mathrm{V}}^{\text{diff}}h^3}{\nu \kappa}  =  \frac{g \alpha Q^{\text{gt}} h^4}{\rho c_p\nu \kappa^2} \approx 1.56\times 10^{24} \per \label{gT17}
\eeq
The numerical estimate in \eqref{gT17} uses $h$ and $\Delta T^{\text{diff}}_{\mathrm{V}}$ above with  $\nu \approx10 \kappa \approx 10^{-6}$ m$^2$ s$^{-1}$, $\alpha \approx 2 \times 10^{-4}$ K$^{-1}$ and $g \approx 10$ m\,s$^{-2}$.

In the body of the paper we use a flux Rayleigh number $\RaV$ based on  buoyancy flux $F$, with units m$^2$ s$^{-3}$, rather than the geothermal heat flux $Q^{\text{gt}}$. The relation between these two fluxes is
\beq
F = \frac{g \alpha Q^{\text{gt}}}{\rho c_p}\per
\label{gT19}
\eeq
Using \eqref{gT19} to eliminate $Q^{\text{gt}}$ from \eqref{gT17} produces the flux Rayleigh number   $\RaV$ in \eqref{RaZ}.

\section{Numerical approach}\label{num_method}
\subsection{Numerical protocol and statistics}
The governing equations (\ref{mom7}--\ref{divu7}) with boundary conditions \eqref{ndBCZ} and no-slip conditions for each boundary, are solved numerically using \href{https://nek5000.mcs.anl.gov/}{Nek5000} \citep{Fischer1997,Deville2002}, which has been used extensively in thermal convection studies \cite[e.g.][]{Scheel2013,rein2023}.
The domain is discretised using up to $\mathrm{E}=2048$ elements which have been refined close to all boundaries to properly resolve viscous and thermal boundary layers.
The velocity is discretised within each element using Lagrange polynomial interpolants based on tensor-product arrays of Gauss–Lobatto–Legendre quadrature points.
The polynomial order $N$ on each element varies between $6$ and $10$ in this study.
We use the $3/2$ rule for dealiasing with extended dealiased polynomial order $3N/2$ to compute nonlinear products.
A third-order time stepping using a mixed explicit-implicit backward difference approach is used.
A summary of the simulations physical and numerical parameters is provided in table \ref{table:1}. 

We initialise all simulations with $\bu=0$ and  $b=0$.
Infinitesimal buoyancy perturbations of amplitude $10^{-6}$ are introduced.
Convection grows during a transient that typically lasts for $t\approx h^2/\kappa$, and which is longer as $\Gamma$ increases.
Once the system has reached a statistically-stationary state, various spatio-temporal averages are computed.
We  define the temporal and volume average operator $\left<\right>$ over the whole fluid domain volume $V$ and over time $t^*$ as
\begin{equation}
\langle A\rangle=\frac{1}{t^* V}\int_{t}^{t+t^*}\!\!\!\int_{V}A(\bx,t) \,\mathrm{d}V\,\mathrm{d}t \ .
\label{avg}
\end{equation}
Typical $t^*$ value range between $100 h^2/\kappa$ and $h^2/\kappa$  for the lowest and the largest Rayleigh numbers respectively.

While most of the results discussed below are obtained using Direct Numerical Simulations (DNS), some extreme cases were only accessible via filtered simulations following the approach described in \cite{fischer}.
To distinguish between DNS and filtered simulations, a viscous dissipation criterion has been used. 
A simulation with polynomial order $N$ is considered to be a DNS when the time and volume averaged viscous dissipation $\varepsilon$ varies by less than $5\%$ when compared with the same simulation but using $N+2$ polynomial order.
A simulation failing to satisfy this criteria is labelled as filtered and numerical stability is ensured by using a $1\%$ filter on the last 2 polynomials \citep{fischer}.
Alternatively, one can use the criterion discussed in \cite{Scheel2013} which compares the isotropic Kolmogorov dissipative scale with the numerical grid size.
For all the DNS simulations presented in this study, the numerical grid size is below the Kolmogorov dissipative scale.
\subsection{Summary of the simulation parameters}\label{appA}
Table \ref{table:1} provide all relevant numerical and physical parameters for the simulations performed using DNS or filtered approaches.
Here, $N$ denotes the polynomial order in both the $x$ and $z$ directions, and $\mathrm{E}$ is the total number of elements.

\begin{scriptsize} 
\setlength{\tabcolsep}{0.8pt} 
\renewcommand{\arraystretch}{1.7} 

\begin{longtable}{ccccccccc}
\caption{Simulations summary (DNS or filtered) according to the physical and numerical parameters.} \\
\noalign{\global\arrayrulewidth=0.01mm}
\multicolumn{1}{c}{$\RaV$} &
\multicolumn{1}{c}{$\RaH$} &
\multicolumn{1}{c}{$\Gamma$} &
\multicolumn{1}{c}{$\RaHtrad$} &
\multicolumn{1}{c}{$Pr$} &
\multicolumn{1}{c}{DNS/filtered} &
\multicolumn{1}{c}{$\mathrm{E}$} &
\multicolumn{1}{c}{$N$} \\
\hline
\endfirsthead

\multicolumn{8}{c}%
{{\bfseries \tablename\ \thetable{} -- continued from previous page}} \\
\noalign{\global\arrayrulewidth=0.01mm}
\multicolumn{1}{c}{$\RaV$} &
\multicolumn{1}{c}{$\RaH$} &
\multicolumn{1}{c}{$\Gamma$} &
\multicolumn{1}{c}{$\RaHtrad$} &
\multicolumn{1}{c}{$Pr$} &
\multicolumn{1}{c}{DNS/filtered} &
\multicolumn{1}{c}{$\mathrm{E}$} &
\multicolumn{1}{c}{$N$} \\
\hline
\endhead

\hline \multicolumn{8}{r}{{Continued on next page}} \\
\endfoot

\hline
\endlastfoot

$0$ & $[10^5,10^6,10^7,10^8]$ & $8$  & $5.12\times[10^7,10^8,10^9,10^{10}]$ & $1$ & DNS & $[512,512,1024,1024]$ & $10$ \\
$10$ & $[1.8,2.643]\times10^2$ & $8$  & $[9.21,13.53]\times10^4$             & $1$ & DNS & $256$ & $6$ \\
$10$ & $2.75\times10^2$        & $12$ & $4.75\times10^5$                      & $1$ & DNS & $256$ & $6$ \\
$10$ & $[3.5,4.6]\times10^2$   & $16$ & $[1.43,1.88]\times10^6$               & $1$ & DNS & $256$ & $6$ \\
$10$ & $[10,6.5,8.77]\times10^2$ & $32$ & $[3.28,2.13,2.87]\times10^7$        & $1$ & DNS & $256$ & $6$ \\
$30$ & $10^3$                  & $32$ & $3.28\times10^7$                      & $1$ & DNS & $256$ & $6$ \\

$100$ & $[10^5,10^6,10^7,10^8]$ & $8$  & $5.12\times[10^7,10^8,10^9,10^{10}]$ & $1$ & DNS & $[512,512,1024,1024]$ & $10$ \\
$100$ & $[6,9.1]\times10^2$     & $8$  & $[3.07,4.66]\times10^5$               & $1$ & DNS & $256$ & $6$ \\
$100$ & $8.5\times10^2$         & $12$ & $1.47\times10^6$                      & $1$ & DNS & $256$ & $6$ \\
$100$ & $[11,15]\times10^2$     & $16$ & $[4.51,6.14]\times10^6$               & $1$ & DNS & $256$ & $6$ \\
$100$ & $[10,20.5,28]\times10^2$& $32$ & $[3.28,6.72,9.18]\times10^7$          & $1$ & DNS & $256$ & $8$ \\
$300$ & $10^3$                  & $32$ & $3.28\times10^7$                      & $1$ & DNS & $256$ & $8$ \\

$10^3$ & $[2.2,3.69]\times10^3$ & $8$  & $[1.13,1.89]\times10^6$               & $1$ & DNS & $256$ & $8$ \\
$10^3$ & $2.75\times10^3$       & $12$ & $4.75\times10^6$                      & $1$ & DNS & $256$ & $8$ \\
$10^3$ & $[3.5,5.2]\times10^3$  & $16$ & $[1.43,2.13]\times10^7$               & $1$ & DNS & $256$ & $8$ \\
$10^3$ & $[1,6.6,9.3]\times10^3$& $32$ & $[3.28,21.6,30.5]\times10^7$          & $1$ & DNS & $256$ & $10$ \\
$10^3$ & \begin{tabular}[c]{@{}c@{}}{[}10,30,100\\ 300,$10^3${]}\end{tabular} & $32$ &
\begin{tabular}[c]{@{}c@{}}{[}$3.28\times10^5$, $9.83\times10^5$\\ $3.28\times10^6$, $9.83\times10^6$, $3.28\times10^7${]}\end{tabular}
& $1$ & DNS & $256$ & $10$ \\

$10^4$ & $[0.95,2.66]\times10^4$ & $8$  & $[4.86,13.6]\times10^6$              & $1$ & DNS & $256$ & $8$ \\
$10^4$ & $1.05\times10^4$        & $12$ & $1.81\times10^7$                     & $1$ & DNS & $256$ & $8$ \\
$10^4$ & $[1.25,2.77]\times10^4$ & $16$ & $[5.12,11.35]\times10^7$             & $1$ & DNS & $256$ & $8$ \\
$10^4$ & $1.7\times10^4$         & $24$ & $2.35\times10^8$                     & $1$ & DNS & $512$ & $8$ \\
$10^4$ & $[2.2,3.75]\times10^4$  & $32$ & $[7.21,12.3]\times10^8$              & $1$ & DNS & $512$ & $10$ \\

$10^5$ & $100$                   & $8$  & $5.12\times10^4$                     & $1$ & DNS & $512$ & $8$ \\
$10^5$ & $[0,0.735,2.5]\times10^5$ & $8$ & $[0,3.76,12.8]\times10^7$            & $1$ & DNS & $512$ & $8$ \\
$10^5$ & $0.75\times10^5$        & $12$ & $1.30\times10^8$                     & $1$ & DNS & $512$ & $8$ \\
$10^5$ & $[0.77,2.48]\times10^5$ & $16$ & $[3.15,10.2]\times10^8$              & $1$ & DNS & $512$ & $8$ \\
$10^5$ & $0.85\times10^5$        & $24$ & $1.18\times10^9$                     & $1$ & DNS & $512$ & $8$ \\
$10^5$ & $[0.95,2.46]\times10^5$ & $32$ & $[3.11,8.06]\times10^9$              & $1$ & DNS & $512$ & $10$ \\

$10^6$ & $100$                   & $8$  & $5.12\times10^4$                     & $1$ & DNS & $512$ & $10$ \\
$10^6$ & $[0,0.6,2.25]\times10^6$ & $8$ & $[0,3.07,11.5]\times10^8$            & $1$ & DNS & $512$ & $10$ \\
$10^6$ & $0.62\times10^6$        & $12$ & $1.07\times10^9$                     & $1$ & DNS & $512$ & $10$ \\
$10^6$ & $[0.61,2.5]\times10^6$  & $16$ & $[2.50,10.24]\times10^9$             & $1$ & DNS & $512$ & $10$ \\
$10^6$ & $0.62\times10^6$        & $24$ & $8.57\times10^9$                     & $1$ & DNS & $1024$ & $10$ \\
$10^6$ & $[0.64,2.45]\times10^6$ & $32$ & $[2.10,8.03]\times10^{10}$           & $1$ & DNS & $1024$ & $10$ \\

$10^7$ & $100$                   & $8$  & $5.12\times10^4$                     & $1$ & DNS & $1024$ & $10$ \\
$10^7$ & $[0,0.335,2.03]\times10^7$ & $8$ & $[0,1.72,10.4]\times10^9$            & $1$ & DNS & $1024$ & $10$ \\
$10^7$ & $0.35\times10^7$        & $12$ & $6.05\times10^9$                     & $1$ & DNS & $1024$ & $10$ \\
$10^7$ & $[0.35,2.05]\times10^7$ & $16$ & $[1.43,8.40]\times10^{10}$           & $1$ & DNS & $1024$ & $10$ \\
$10^7$ & $0.35\times10^7$        & $24$ & $4.84\times10^{10}$                  & $1$ & DNS & $1024$ & $10$ \\
$10^7$ & $[0.1,0.35,2]\times10^7$ & $32$ & $[3.28,11.5,65.5]\times10^{10}$      & $1$ & DNS & $1024$ & $10$ \\

$10^8$ & $100$                   & $8$  & $5.12\times10^4$                     & $1$ & DNS & $1024$ & $10$ \\
$10^8$ & $[0,0.2,1.76]\times10^8$ & $8$ & $[0,1.02,9.01]\times10^{10}$         & $1$ & DNS & $1024$ & $10$ \\
$10^8$ & $0.2\times10^8$         & $12$ & $3.46\times10^{10}$                  & $1$ & DNS & $1024$ & $10$ \\
$10^8$ & $[0.2,1.7]\times10^8$   & $16$ & $[8.19,69.6]\times10^{10}$           & $1$ & DNS & $1024$ & $10$ \\
$10^8$ & $0.2\times10^8$         & $24$ & $2.76\times10^{11}$                  & $1$ & DNS & $2048$ & $10$ \\
$10^8$ & $[0.21,1.7]\times10^8$  & $32$ & $[6.88,55.7]\times10^{11}$           & $1$ & DNS & $2048$ & $10$ \\

$10^9$ & $100$                   & $8$  & $5.12\times10^4$                     & $1$ & filtered & $2048$ & $10$ \\
$10^9$ & $[0,0.14,1.7]\times10^9$ & $8$ & $[0,7.17,87.0]\times10^{10}$         & $1$ & filtered & $2048$ & $10$ \\
$10^{10}$ & $9\times10^{8}$      & $8$  & $4.61\times10^{11}$                  & $1$ & filtered & $2048$ & $10$ \\

$10^7$ & $3.35\times10^6$ & 8 & $1.71\times10^9$& \begin{tabular}[c]{@{}c@{}}{[}2,3,4,6,7,8\\10,11,12,13{]}\end{tabular}& DNS & $1024$ & $10$ \\

$10^7$ & $[3.19,2.02]\times10^6$ & $8$  & $[1.63,1.03]\times10^9$              & $4$ & DNS & $1024$ & $10$ \\
$10^7$ & $[3.15,1.96]\times10^6$ & $8$  & $[1.61,1.00]\times10^9$             & $7$ & DNS & $1024$ & $10$ \\
$10^7$ & $[3.1,1.94]\times10^6$  & $8$ & $[1.61,0.99]\times10^9$              & $10$ & DNS & $1024$ & $10$ \\
$10^7$ & $[3.07,1.93]\times10^6$ & $8$ & $[1.57,0.98]\times10^9$              & $13$ & DNS & $1024$ & $10$ \\

\label{table:1}
\end{longtable} 
\end{scriptsize}


\section{A large aspect ratio  regime}\label{appLowRa}
We consider the non-dimensional governing equations \eqref{mom7}--\eqref{divu7} with the boundary conditions \eqref{ndBCZ}.
We use the notation $\ep = \Gamma^{-1} \ll 1$ and focus on the two-dimensional case with    a  streamfunction $\psi$ such that $(u,w)=(-\psi_z,\psi_x)$. We eliminate  the pressure to form the out-of-plane vorticity equation and define the rescaled horizontal coordinate $X=\epsilon x$ and   rescaled buoyancy and Rayleigh numbers
\begin{equation}
    B=\epsilon b, \qquad \hatRaF= \epsilon\RaV, \qquad \hatRaH= \epsilon\RaH.
\end{equation}
After these machinations, the steady equations of motion are
\begin{align}
     \Pr^{-1}\ep \big[\psi_X(\ep^2 \psi_{XX} + \psi_{zz})_z  &- \psi_z(\ep^2 \psi_{XX} + \psi_{zz})_X\big] \nonumber \\  &= B_X +  \ep^4\psi_{XXXX} + 2 \epsilon^2 \psi_{XXzz} +  \psi_{zzzz},
    \label{asvor1}\\
    \epsilon \big[\psi_X B_z - \psi_z B_X\big] &= \epsilon^2B_{XX} + B_{zz},
    \label{asnrj1}
\end{align}
with boundary conditions
\beq
B(X,1) = - \hatRaH \tau(X) \com \qquad B_z(X,0) = - \hatRaF\per
\label{bcd}
\eeq
Above, $\tau(X) = \cos (2 \pi X)$ and $-1/2<X < 1/2$.

\begin{figure}
  \centering
\includegraphics[width=1.0\textwidth]{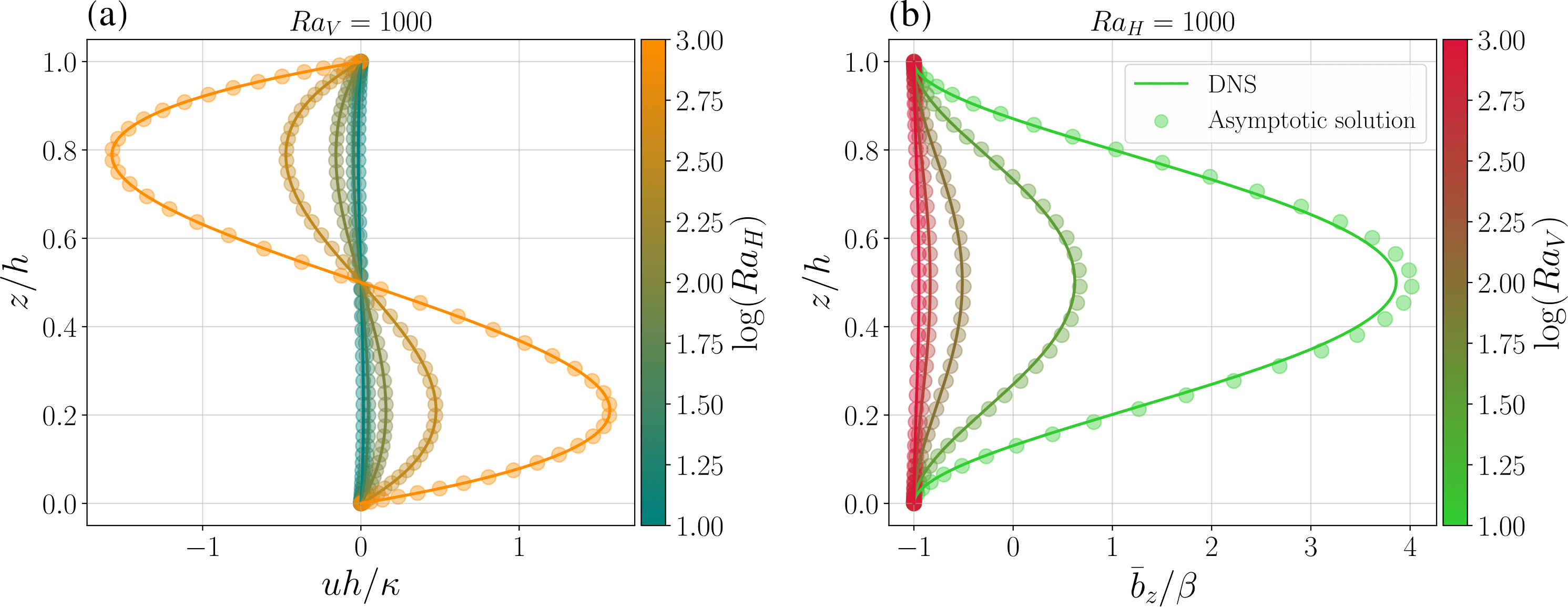}
    \caption{$(a)$ Vertical profile of the horizontal velocity at $X=1/2$ for $\RaH \in [10,30,100,300,1000]$ for fixed $\RaV=1000$. $(b)$ Vertical profile of $\bar b_z$ for $\RaV \in [10,30,100,300,1000]$ for a fix $\RaH=1000$.
    The continuous lines (--) represent the DNS data and circles ($\circ$) represent the asymptotic solution (\eqref{psi0} for $(a)$ and based on \eqref{B0},\eqref{B1} for $(b)$ ). The input parameters are $\Gamma=32$ and $Pr=1.0$. }
  \label{fig11}
\end{figure}

With small $\ep$ we can develop a lubrication-type approximation by considering the distinguished limit in which $\ep \to 0$ with $\hatRaH$ and $\hatRaF$ fixed and order unity.
Our objective is to compute the two-dimensional steady-state solution for Rayleigh numbers sufficiently small to remain below the threshold for any instability.
We seek a solution of \eqref{asvor1} and \eqref{asnrj1} for the streamfunction and the rescaled buoyancy expanded in powers of $\epsilon$ as
\begin{equation}
    \begin{aligned}
        B(X,z)&= B_0(X,z)+ \epsilon B_1(X,z)  + \cdots,\\
        \psi(X,z) &= \psi_0(X,z) + \epsilon \psi_1(X,z) + \cdots~~.
    \end{aligned}
\end{equation}
Substituting these expansions into \eqref{asvor1} and \eqref{asnrj1} gives at leading order
\begin{equation}
    {B_0}_{zz} = 0\qquad \qquad \text{and} \qquad \qquad {\psi_0}_{zzzz} = -{B_0}_X.
     \label{0psi}
\end{equation}
The solution of the buoyancy equation satisfying the the top and bottom boundary conditions \eqref{bcd} is
\begin{equation}
    B_0(X,z) = \hatRaF\big(1-z\big) - \hatRaH~\tau(X)\per
    \label{B0}
\end{equation}
We can then solve the vorticity equation in  \eqref{0psi} to find 
\begin{equation}
    \psi_0(X,z) = \hatRaH \P(z)\tau_X,\qquad \text{with} \qquad \P(z)=\frac{z^2(1-z)^2}{24}.
    \label{psi0}
\end{equation}
The polynomial $\P(z)$ is defined by $\P^{(\mathrm{IV})}=1$, with $\P(0)=\P^{\prime}(0)=\P(1)=\P^{\prime}(1)=0$.
The leading-order  buoyancy field, $B_0(X,z)$ in \eqref{B0}, is unaffected by the flow, i.e.~there is no feedback from the velocity field on the buoyancy profile.
To capture the first-order correction due to advection, we analyze the buoyancy equation at order $O(\epsilon)$,
\begin{equation}
    {B_1}_{zz} = {\psi_0}_X{B_0}_z - {\psi_0}_z{B_0}_X =(\psi_0{B_0}_z)_X - (\psi_0{B_0}_X)_z  .
    \label{B1zz}
\end{equation}
Integrating \eqref{B1zz} twice in $z$ gives
\begin{equation}
    \begin{aligned}
            B_1= -\hatRaF \hatRaH \tau_{XX}\Q_0(z) + \hatRaH{}^{\!2} \tau_X^2\Q_1(z),
     \label{B1}
    \end{aligned}
\end{equation}
with 
\begin{align}
    \Q_0(z) &= \frac{z^6}{720}-\frac{z^5}{240}+\frac{z^4}{288}-\frac{1}{1440} \com \\  \Q_1(z) &= \frac{z^5}{120}-\frac{z^4}{48}+\frac{z^3}{72}-\frac{1}{720}.
\end{align}
Polynomials $\Q_0(z)$ and $\Q_1(z)$ are defined by $\Q_0^{\prime \prime } = \P$ and $\Q_1^{\prime} = \P$, with $\Q_{1}(1)=\Q_{0}(1)=\Q_{0}^{\prime}(0)=0$, satisfying the buoyancy boundary conditions.

\autoref{fig11} compares the vertical profiles of $\bar b_z$ and the horizontal velocity obtained from a DNS with the approximate asymptotic solution derived earlier for several $\RaH$ at $\RaV$ ranging from 1 to 1000.
The asymptotic solution neglects sidewall effects. Thus we use a large aspect ratio, $\Gamma = 32$, to minimize the influence of lateral boundaries in the bulk region.
Overall, the asymptotic solution shows good agreement with the DNS results for both the buoyancy gradient and velocity fields.

\autoref{fig11}(b) shows that when $\RaV \approx \RaH$ the flow is negatively stratified throughout the depth. Conversely, when $\RaV \ll \RaH$ the bulk of the layer exhibits positive stratification.  
For fixed $\RaV$, increasing $\RaH$ strengthens this stable stratification and extends it over a larger fraction of the depth, so that the domain as a whole can become positively stratified in a volume‐mean sense.  
The condition at which the system first achieves positive mean stratification can be predicted from the lubrication solution derived above.


\begin{figure}
  \centering
  \includegraphics[width=0.9\textwidth]{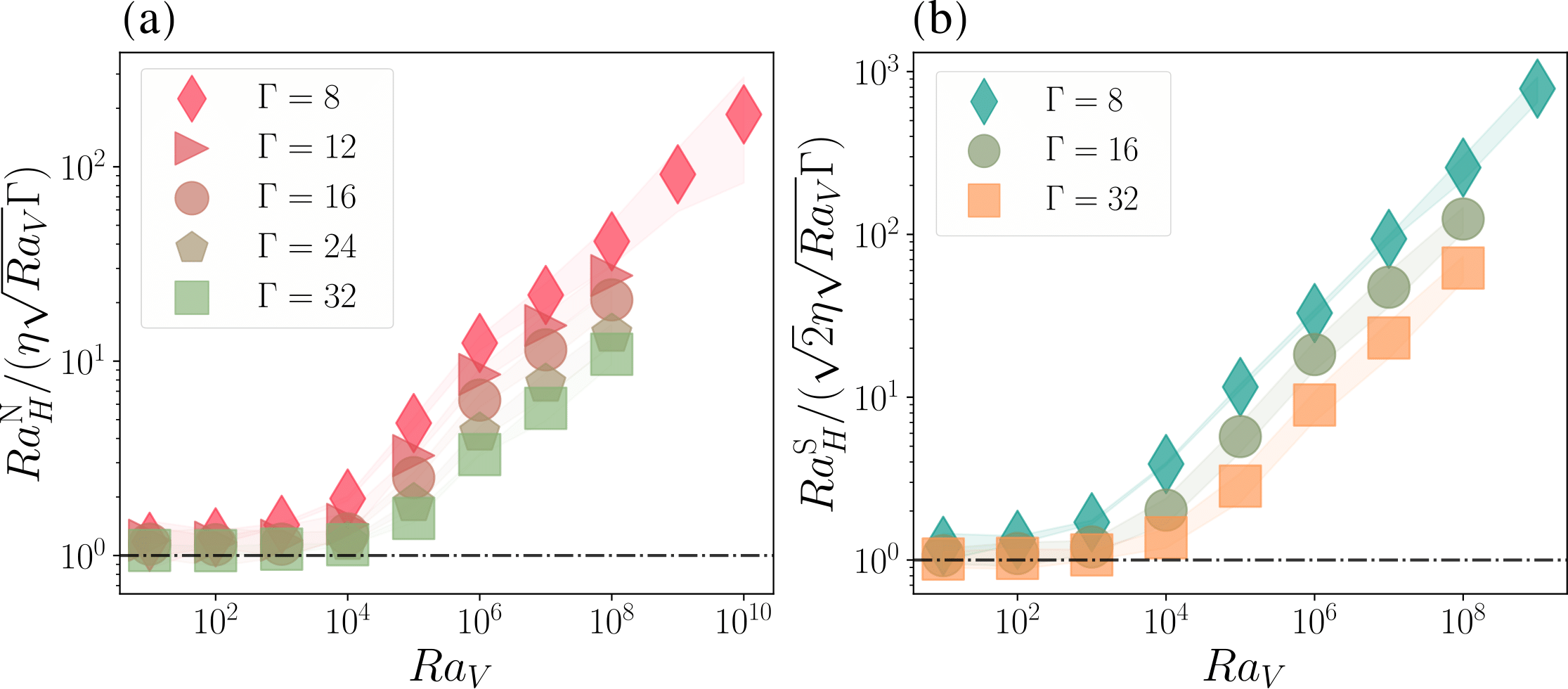}
    \caption{Neutral and strong horizontal Rayleigh numbers, $(a)$ $\RaHN$ and $(b)$ $\RaHstrg$,  
compensated by their respective theoretical scalings in \eqref{RaHN_low} and plotted
against the vertical Rayleigh number $\RaV$ for several aspect ratios~$\Gamma$.
}
  \label{fig12}
\end{figure}

Using equations \eqref{B0} and \eqref{B1} and taking derivatives with respect to $z$, the vertical buoyancy gradient can be expressed as
\begin{equation}
b_z = -\RaV + \epsilon^2 \Big[ -\RaV\RaH\tau_{XX}\Q_0^\prime + \RaH{\!^2} \tau_X^2 \Q_1^\prime \Big].
\label{bz}
\end{equation}
The global average of \eqref{bz} is
\begin{equation}
\langle b_z \rangle = -\RaV + \bigg(\frac{\epsilon \RaH}{\eta}\bigg)^2,
\label{gb1z}
\end{equation}
where $\eta^2 = -1/(\Q_1(0)\overline{\tau_X^2}) \approx 36.5$ is a numerical constant resulting from the global averaging of the vertical and horizontal profiles of \eqref{bz}.
Examining the condition under which \eqref{gb1z} vanishes and/or reverse the imposed bottom flux provides insight into the scaling of neutral and strong stratification in the low Rayleigh regime.
Setting $\langle b_z \rangle = 0$  and  $\langle b_z \rangle = \RaV$ gives respectively,
\begin{equation}
\RaHN = \eta \sqrt{\RaV}  \Gamma \quad\mbox{and}\quad \RaHstrg = \sqrt{2}\eta \sqrt{\RaV}  \Gamma.
\label{RaHN_low}
\end{equation}
In \autoref{fig12}, we show $\RaHN$ and $\RaHstrg$ compensated by their respective scaling relations \eqref{RaHN_low}, plotted as a function of $\RaV$. The results exhibit consistent trends, confirming the predicted $\RaV$- and $\Gamma$-dependence in the low-Rayleigh-number regime, i.e.~for $\RaV$ and $\RaH \lesssim 10^3$. Furthermore, the relation \eqref{RaHN_low} also captures the prefactor with reasonable accuracy, although a slight deviation from the data is noticeable. Improving the quantitative agreement would require evaluating the $O(\epsilon^2)$ correction to the buoyancy field. We conclude by noting that \eqref{RaHN_low} does not involve $\Pr$.

\bigskip
\bigskip

\noindent

\bibliographystyle{jfm}
\bibliography{convec2.bib}

@article{couston2022,
  title={Competition between {R}ayleigh--{B}{\'e}nard and horizontal convection},
  author={Couston, L.-A. and Nandaha, J. and Favier, B.},
  journal={J.~Fluid Mech.},
  volume={947},
  pages={A13},
  year={2022},
  publisher={Cambridge University Press}
}

@article{deLavergne2016,
  title={On the consumption of {A}ntarctic {B}ottom {W}ater in the abyssal ocean},
  author={De Lavergne, C. and Madec, G. and Le Sommer, J. and Nurser, A. J. G. and Naveira Garabato, A. C},
  journal={J.~Phys. Oceanogr.},
  volume={46},
  pages={635--661},
  year={2016}
}

@article{emile2009,
  title={Geothermal heating, diapycnal mixing and the abyssal circulation},
  author={Emile-Geay, J. and Madec, G.},
  journal={Ocean Science},
  volume={5},
  pages={203--217},
  year={2009},
  publisher={Copernicus Publications G{\"o}ttingen, Germany}
}

@article{paparella2002,
	Author = {Paparella, F. and Young, W. R.},
	Doi = {10.1017/S0022112002001313},
	Journal = {J.~Fluid Mech.},
	Pages = {205--214},
	Title = {Horizontal convection is non-turbulent},
	Volume = {466},
	Year = {2002},
}

@article{munk1998,
	Author = {Munk, W. and Wunsch, C. I.},
	Issn = {0967-0637},
	Journal = {Deep Sea Res. Part {I}},
	Pages = {1977--2010},
	Title = {Abyssal recipes {II}: energetics of tidal and wind mixing},
	Volume = {45},
	Year = {1998}}

@article{wunsch2016,
  title={Tides of global ice-covered oceans},
  author={Wunsch, C.},
  journal={Icarus},
  volume={274},
  pages={122--130},
  year={2016},
  publisher={Elsevier}
}

@article{soderlund2019,
  title={Ocean dynamics of outer solar system satellites},
  author={Soderlund, K. M.},
  journal={Geophys. Res. Lett.},
  volume={46},
  pages={8700--8710},
  year={2019},
  publisher={Wiley Online Library}
}

@article{rossby1965,
  title={On thermal convection driven by non-uniform heating from below: an experimental study},
  author={Rossby, H. T.},
  journal={Deep Sea Res.},
  volume={12},
  pages={9--16},
  year={1965}  
}

@book{goluskin2016,
  title={Internally heated convection and Rayleigh-B{\'e}nard convection},
  author={Goluskin, D.},
  year={2016},
  publisher={Springer},
  address={Cham, Switzerland}
}

@article{rocha2020,
  title={The {N}usselt numbers of horizontal convection},
  author={Rocha, C.B. and Constantinou, N.C. and Llewellyn Smith, S. G.  and Young, W.R.},
  journal={Journal of Fluid Mechanics},
  volume={894},
  pages={A24},
  year={2020},
  publisher={Cambridge University Press}
}

@article{mullarney2006,
  title={The effects of geothermal heating on the ocean overturning circulation},
  author={Mullarney, J. C. and Griffiths, R. W. and Hughes, G. O.},
  journal={Geophys. Res. Lett.},
  volume={33},
  year={2006},
  publisher={Wiley Online Library}
}

@article{wang2016,
  title={Laboratory simulation of the geothermal heating effects on ocean overturning circulation},
  author={Wang, F. and Huang, S.-D. and Zhou, S.-Q. and Xia, K.-Q.},
  journal={J.~Geophys. Res.: Oceans},
  volume={121},
  pages={7589--7598},
  year={2016},
  publisher={Wiley Online Library}
}

@article{pierrehumbert2011,
  title={Climate of the {N}eoproterozoic},
  author={Pierrehumbert, R. T. and Abbot, D. S. and Voigt, A. and Koll, D.},
  journal={Ann. Rev. Earth Planetary Sci.},
  volume={39},
  pages={417--460},
  year={2011},
  publisher={Annual Reviews}
}

@article{hoffman2002,
  title={The snowball {Earth} hypothesis: testing the limits of global change},
  author={Hoffman, P. F and Schrag, D. P},
  journal={Terra Nova},
  volume={14},
  pages={129--155},
  year={2002},
  publisher={Wiley Online Library}
}

@article{jansen2016,
  title={The turbulent circulation of a {S}nowball {E}arth ocean},
  author={Jansen, M. F.},
  journal={J.~Phys. Oceanogr.},
  volume={46},
  pages={1917--1933},
  year={2016}
}

@article{ashkenazy2016,
  title={Variability, instabilities, and eddies in a Snowball Ocean},
  author={Ashkenazy, Y. and Tziperman, E.},
  journal={Journal of Climate},
  volume={29},
  pages={869--888},
  year={2016}
}

@article{bire2022,
  title={Exploring ocean circulation on icy moons heated from below},
  author={Bire, S. and Kang, W. and Ramadhan, A. and Campin, J.-M. and Marshall, J.},
  journal={J.~Geophys. Res.: Planets},
  volume={127},
  pages={e2021JE007025},
  year={2022},
  publisher={Wiley Online Library}
}

@article{wunsch2004,
  title={Vertical mixing, energy, and the general circulation of the oceans},
  author={Wunsch, C. and Ferrari, R.},
  journal={Annu. Rev. Fluid Mech.},
  volume={36},
  pages={281--314},
  year={2004},
  publisher={Annual Reviews}
}

@article{Scheel2013,
   title = {Resolving the fine-scale structure in turbulent {R}ayleigh-{B}\'enard convection},
   author = {J. D. Scheel and M. S. Emran and J. Schumacher},
   journal = {New J.~Phys.},
   pages = {113063},
   volume = {15},
   year = {2013},
   publisher = {IOP Publishing},
}

@article{rein2023, 
title={Interaction between forced and natural convection in a thin cylindrical fluid layer at low {P}randtl number},  
journal={J.~ Fluid Mech.},
volume={977},
author={Rein, F. and Car\'enini, L. and Fichot, F. and Favier, B. and Le Bars, M.},
year={2023},
pages={A26}
}

@article{Fischer1997,
   author = {P. F. Fischer},
   journal = {J.~Comp. Phys.},
   pages = {84--101},
   title = {An Overlapping {S}chwarz Method for Spectral Element Solution of the Incompressible {N}avier--{S}tokes Equations},
   volume = {133},
   year = {1997},
}

@book{Deville2002,
   author = {M. O. Deville and P. F. Fischer and E. H. Mund},
   city = {Cambridge},
   doi = {10.1017/CBO9780511546792},
   publisher = {Cambridge University Press},
   title = {High-Order Methods for Incompressible Fluid Flow},
   year = {2002},
   address = {Cambridge, UK}
}

@Article{FISCHER,
    title = {Filter-based stabilization of spectral element methods},
    journal = {C.~R.~Acad. Sci. I Math.},
    volume = {332},
    pages = {265--270},
    year = {2001},
    author = {P. Fischer and  J. Mullen },
}

@article{zhang2024ocean,
  title={Ocean weather systems on icy moons, with application to {E}nceladus},
  author={Zhang, Y. and Kang, W. and Marshall, J.},
  journal={Science Advances},
  volume={10},
  pages={6857},
  year={2024},
  publisher={American Association for the Advancement of Science}
}

@article{GL2000, 
title={Scaling in thermal convection: a unifying theory}, 
volume={407}, 
journal={J.~Fluid Mech.},
publisher={Cambridge University Press}, 
author={Grossmann, S. and Lohse, D.}, 
year={2000}, 
}

@article{ANRHC2008,
   author = "Hughes, G. O. and Griffiths, R. W.",
   title = "Horizontal Convection", 
   journal= "Annu. Rev. Fluid Mech.",
   year = "2008",
   volume = "40",
   pages = "185--208",
   publisher = "Annual Reviews",
  }

@article{rossby1998,
  title={Numerical experiments with a fluid heated non-uniformly from below},
  author={Rossby, H. T.},
  journal={Tellus A},
  volume={50},
  pages={242--257},
  year={1998},
  publisher={Taylor \& Francis}
}

@article{THOMA2009,
title = {Modelling flow and accreted ice in subglacial {L}ake {C}oncordia, {A}ntarctica},
journal = {Earth Planetary Sci. Lett.},
volume = {286},
pages = {278--284},
year = {2009},
author = {M. Thoma and K. Grosfeld and I. Filina and C. Mayer},
}

@article{couston2021,
author = {L.-A. Couston  and M. Siegert },
title = {Dynamic flows create potentially habitable conditions in {Antarctic} subglacial lakes},
journal = {Science Advances},
volume = {7},
pages = {eabc3972},
year = {2021},
}

@article{johnston2009,
  title = {Comparison of Turbulent Thermal Convection between Conditions of Constant Temperature and Constant Flux},
  author = {Johnston, H. and Doering, C. R.},
  journal = {Phys. Rev. Lett.},
  volume = {102},
  pages = {064501},
  numpages = {4},
  year = {2009},
  month = {Feb},
  publisher = {American Physical Society},
}

@article{otero2002, 
title={Bounds on {Rayleigh--B\'enard} convection with an imposed heat flux}, 
volume={473}, 
journal={Journal of Fluid Mechanics}, 
author={Otero, J. and Wittengerb, R. W. and Worthing, R. A. and Doering, C. R.}, 
year={2002}, 
pages={191--199}}

@article{Lemasquerier2023,
author = {Lemasquerier, D. G. and Bierson, C. J. and Soderlund, K. M.},
title = {Europa's Ocean Translates Interior Tidal Heating Patterns to the Ice-Ocean Boundary},
journal = {AGU Advances},
volume = {4},
pages = {e2023AV000994},
year = {2023}
}

@article{gerkema2008,
  title={Geophysical and astrophysical fluid dynamics beyond the traditional approximation},
  author={Gerkema, T. and Zimmerman, J. T. F. and Maas, L. R. M and Van Haren, H.},
  journal={Rev. Geophys.},
  volume={46},
  page={2006RG000220},
  year={2008},
  publisher={Wiley Online Library}
}

@article{winters2009,
  title={Available potential energy and buoyancy variance in horizontal convection},
  author={Winters, K.B. and Young, W. R.},
  journal={Journal of Fluid Mechanics},
  volume={629},
  pages={221--230},
  year={2009},
  publisher={Cambridge University Press}
}

@article{GL2001,
  title = {Thermal Convection for Large {P}randtl Numbers},
  author = {Grossmann, S. and Lohse, D.},
  journal = {Phys. Rev. Lett.},
  volume = {86},
  pages = {3316--3319},
  year = {2001},
  publisher = {American Physical Society},
}

@article{Shishkina2016,
author = {Shishkina, O. and Grossmann, S. and Lohse, D.},
title = {Heat and momentum transport scalings in horizontal convection},
journal = {Geophys. Res. Lett.},
volume = {43},
number = {3},
pages = {1219-1225},
year = {2016}
}

@article{lindborg2023scaling,
  title={Scaling in {R}ayleigh--{B}{\'e}nard convection},
  author={Lindborg, Erik},
  journal={Journal of Fluid Mechanics},
  volume={956},
  pages={A34},
  year={2023},
  publisher={Cambridge University Press}
}

@article{doering2020turning,
  title={Turning up the heat in turbulent thermal convection},
  author={Doering, Charles R},
  journal={Proceedings of the National Academy of Sciences},
  volume={117},
  number={18},
  pages={9671--9673},
  year={2020},
  publisher={National Academy of Sciences}
}

@article{Passaggia_Cohen_Scotti_2024,
title={Limiting regimes of turbulent horizontal convection. Part 2. Large Prandtl numbers},
volume={997},
journal={Journal of Fluid Mechanics},
author={Passaggia, PY and Cohen, N. F. and Scotti, A.},
year={2024},
pages={A6}}

@article{Passaggia_Scotti_2024,
title={Limiting regimes of turbulent horizontal convection. Part 1. Intermediate and low Prandtl numbers},
volume={997},
journal={Journal of Fluid Mechanics}, 
author={Passaggia, PY and Scotti, A.}, 
year={2024}, 
pages={A5}}
\end{document}